\documentclass[lettersize,journal]{IEEEtran}
\usepackage{amsmath,amsfonts}
\usepackage[linesnumbered,ruled,vlined]{algorithm2e}
\usepackage{array}
\usepackage[caption=false,font=footnotesize,labelfont=rm,textfont=rm]{subfig}
\usepackage{textcomp}
\usepackage{stfloats}
\usepackage{url}
\usepackage{verbatim}
\usepackage{graphicx}
\usepackage{cite}

\usepackage{caption}
\usepackage{booktabs}
\usepackage{xspace}
\usepackage[T1]{fontenc}
\usepackage{xcolor}
\usepackage{multirow}
\usepackage{bbding}
\usepackage{newtxtext}
\usepackage{pifont}

\newcommand{\tool}{\textsc{SWA-LDM}\xspace}

\makeatletter
\DeclareRobustCommand\onedot{\futurelet\@let@token\@onedot}
\def\@onedot{\ifx\@let@token.\else.\null\fi\xspace}

\def\eg{\emph{e.g}\onedot} 

\def\ie{\emph{i.e}\onedot}

\def\vs{\emph{vs}\onedot}

\makeatother

\begin{document}

% \title{A Sample Article Using IEEEtran.cls\\ for IEEE Journals and Transactions}
\title{\emph{\tool}: Toward Stealthy Watermarks for Latent Diffusion Models}

\author{Zhonghao Yang%~\IEEEmembership{Member,~IEEE}
, Linye Lyu, Xuanhang Chang, Daojing He, Cheng Zhuo, Yu LI
%,~\IEEEmembership{Life Fellow,~IEEE}
        % <-this % stops a space
        % <-this % stops a space
\thanks{Zhonghao Yang and Daojing He are with the Software Engineering Institute, East China Normal University, 200062, China. (e-mail: ashzhonghao@gmail.com; hedaojinghit@163.com)}
\thanks{Linye Lyu and Xuanhang Chang are with the School of Computer Science and Technology, Harbin Institute of Technology (Shen Zhen), 518055, China.}
\thanks{Cheng Zhuo and Yu Li are with College of Integrated Circuits, Zhejiang University, Hangzhou, 310000, Zhejiang, China (e-mail: yu.li.sallylee@gmail.com).}
}


\IEEEpubid{0000--0000/00\$00.00~\copyright~2021 IEEE}

\maketitle

\begin{abstract}
Latent Diffusion Models (LDMs) have established themselves as powerful tools in the rapidly evolving field of image generation, capable of producing highly realistic images. However, their widespread adoption raises critical concerns about copyright infringement and the misuse of generated content. Watermarking techniques have emerged as a promising solution, enabling copyright identification and misuse tracing through imperceptible markers embedded in generated images. Among these, latent-based watermarking techniques are particularly promising, as they embed watermarks directly into the latent noise without altering the underlying LDM architecture.

In this work, we demonstrate—for the first time—that such latent-based watermarks are practically vulnerable to detection and compromise through systematic analysis of output images' statistical patterns. To counter this, we propose SWA-LDM (Stealthy Watermark for LDM), a lightweight framework that enhances stealth by dynamically randomizing the embedded watermarks using the Gaussian-distributed latent noise inherent to diffusion models.
By embedding unique, pattern-free signatures per image, SWA-LDM eliminates detectable artifacts while preserving image quality and extraction robustness. Experiments demonstrate an average of 20\% improvement in stealth over state-of-the-art methods, enabling secure deployment of watermarked generative AI in real-world applications.
\end{abstract}


\section{Introduction}
\label{sec:intro}
\IEEEPARstart{T}{he} Latent Diffusion Models (LDMs)  \cite{rombach2022high} represent a significant advancement in efficient, high-quality image generation.
By leveraging Variational Autoencoders (VAEs) \cite{Kingma2014}, LDMs transfer diffusion model operations from pixel space to latent space, allowing U-Net \cite{ronneberger2015u} architectures to perform denoising in a lower-dimensional space. 
This shift dramatically enhances computational efficiency, enabling companies and individuals with limited resources to train models for commercial usage.
Consequently, popular models such as Stable Diffusion (SD) \cite{rombach2022high}, DALL-E 2 \cite{ramesh2022hierarchical}, and Midjourney \cite{midjourney} have emerged, facilitating the generation of high-quality, realistic images via user-accessible APIs.

The rapid advancements of LDMs have introduced critical challenges, particularly concerning copyright infringement and the potential misuse of generated content. Copyright violations arise when malicious actors steal and resell proprietary diffusion models/images, resulting in substantial financial losses for original creators. Additionally, the capability to generate hyper-realistic images has been exploited by individuals disseminating misinformation and fake news, thereby undermining public trust and social stability. Addressing these issues is paramount for safeguarding intellectual property rights and maintaining societal integrity.

\begin{figure}
    \centering
    \includegraphics[width=\linewidth]{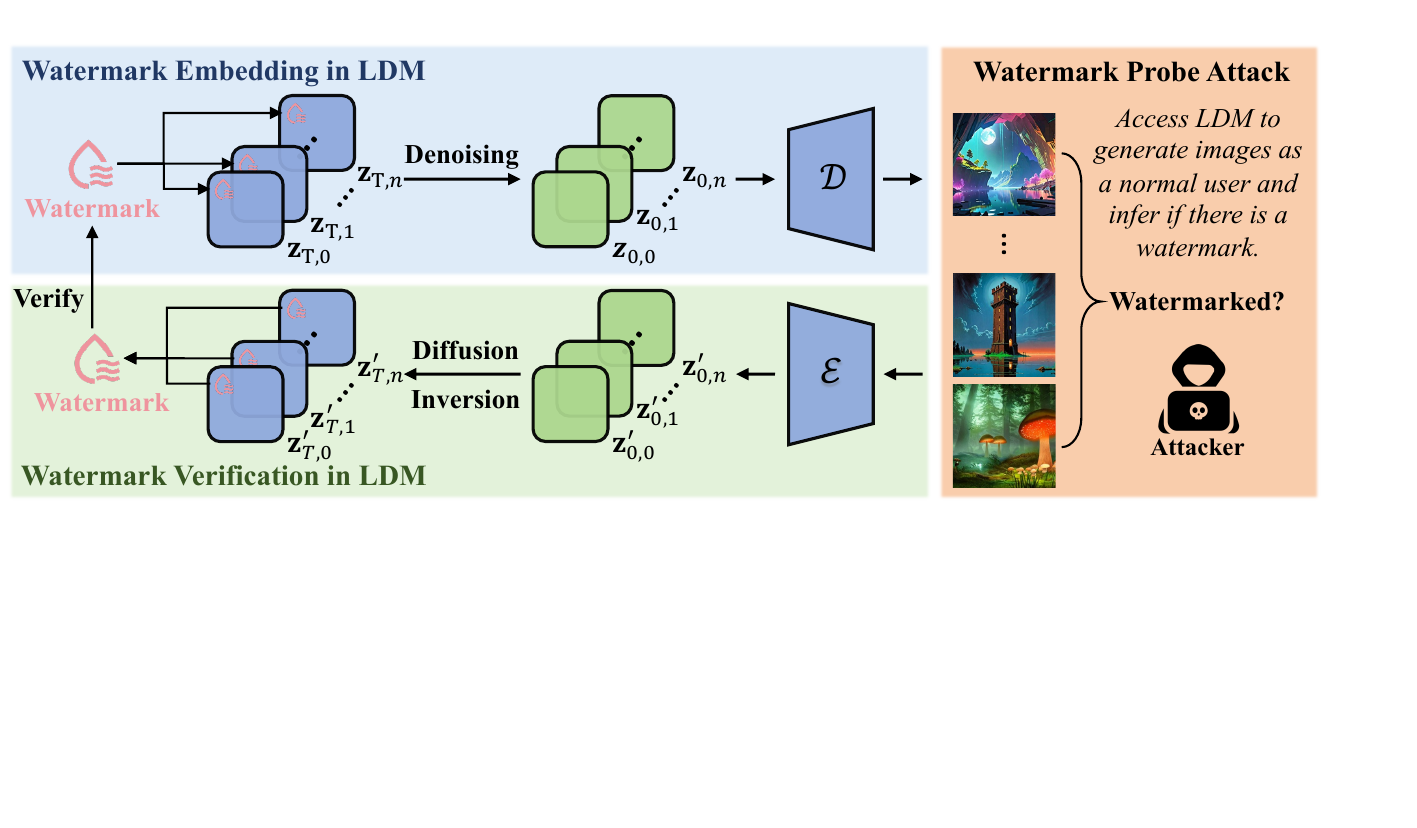}
    % \vspace{-0.6cm}
    \caption{
    The pipeline of the latent-based watermarking method. 
    % The watermark is embedded into the latent noise, which is subsequently fed into the LDM for image generation.
    % They often add the same watermark signal to different generated images, which attackers can exploit to detect the presence of watermark. 
    % This stealthiness vulnerability can facilitate watermark removal, causing the risk of copyright infringement.  
    }
    % \vspace{-0.6cm}
    \label{fig:l-r-b_method_framework}
\end{figure}

To address these challenges, modern LDMs embed imperceptible watermarks during generation, which can later be extracted via specialized secret decoding methods to authenticate image origins. 
Unlike the traditional post-processing watermarking that add watermarks \textit{after} LDMs have generated the images ~\cite{Ruanaidh_Dowling_Boland_1996,O’Ruanaidh_Pun_2002,cox2007digital,zhang2019robust} , emerging in-generation-process approaches~\cite{fernandez2023stable, wen2023tree, yang2024gaussian, lei2024diffusetrace, feng2024aqualora} integrate watermarks directly into the image generation process for better image fidelity.
These methods can be divided into two types: model-based~\cite{fernandez2023stable, feng2024aqualora} and latent-based methods~\cite{wen2023tree, yang2024gaussian, lei2024diffusetrace}. 
The former embeds watermarks by modifying LDMs' architectural component, such as VAE or U-Net. In contrast, the latent-based methods, embed the watermark to the latent noise \textit{prior} to the denoising process, as illustrated in Fig.~\ref{fig:l-r-b_method_framework}. 
Latent-based methods show great promise by avoiding model retraining, maintaining image quality, and offering plug-and-play usability—making them ideal for practical deployment.

However, we demonstrate—for the first time—that the stealthiness of these latent-based watermarks can be \textit{practically} compromised. 
We identify a critical flaw: existing methods often embed identical watermarks across multiple distinct generated images, \ie, for each watermark, they embed the same pattern for multiple outputs. These recurring signals act as detectable fingerprints of the watermark. 
Leveraging this insight, we design a high-accuracy \textbf{watermark probe attack} that detects these patterns under minimal attacker assumptions: no prior watermarked images are required, and the attack succeeds with only black-box API access to the target LDM, coupled statistical analysis of its outputs. Such practicality highlights severe risks to real-world systems relying on such watermarks.

\IEEEpubidadjcol
To mitigate this risk, we present \textbf{\tool} – a lightweight framework designed to augment existing latent-based watermarking methods with enhanced stealth capabilities. At its core, \tool introduces \textit{dynamic watermark randomization} through a seed channel derived from Gaussian-distributed latent noise, enabling unique fingerprints for each generated image. This approach fundamentally disrupts pattern recognition in probe attacks by eliminating static watermark signatures.
Since such a seed channel is a subset of the latent variable used for image generation, it requires no extra store or management, making it lightweight. 
During watermark verification, \tool reconstructs the latent variable via diffusion inversion and extracts the seed to retrieve the original watermark. However, inaccuracies may arise due to diffusion inversion and image transmission noises. To mitigate this, we propose an enhancement algorithm to store redundant seeds in the seed channel while preserving its distribution. This dual strategy of dynamic watermarking and noise-resilient encoding achieves both stealth and robustness.

Our contributions are summarized as follows:

\begin{itemize}
    \item 
    We propose the first \textit{practical} watermark probe attack against latent-based LDM methods, enabling detection of watermarks without the secret watermark decoders or other prior knowledge of the watermark—revealing critical stealth limitations in current strategies.
    \item  
    We introduce \tool, a plug-and-play solution that leverages latent noise randomness to generate image-specific watermarks, eliminating detectable patterns with zero management overhead.
    \item 
    We develop a distribution-preserving redundant seed encoding algorithm, improving watermark extraction accuracy under noise while maintaining latent space integrity.
\end{itemize}
Extensive experimental results show that \tool effectively improves the stealthiness of latent-based watermarks while achieving competitive visual quality, image-text similarity, and watermarking robustness.

\section{Background and Related Works}
\textbf{Latent Diffusion Models.} 
Latent diffusion models are a computationally efficient version of diffusion models \cite{rombach2022high}. LDMs leverage a pretrained autoencoder to compress image $\mathbf{x}_0$ in RGB space into a lower dimensional latent representation $\mathbf{z}_0 $, which significantly reduces the computational complexity of LDMs' training and sampling. More specifically, during training, the encoder $\mathcal{E}$ encodes the image $ \mathbf{x}_0 $ into a latent representation by $\mathbf{z}_0 = \mathcal{E}(\mathbf{x}_0)$. Then, LDMs conduct both diffusion and denoising process in the latent space. The diffusion process converts $\mathbf{z}_0$ to a latent noise $\mathbf{z}_T$ and the denoising process recovers the image latent $\tilde{\mathbf{z}}_0$  from $\mathbf{z}_T$ over  $T$ timesteps. Afterwards, the decoder $\mathcal{D}$ reconstructs the image $\tilde{\mathbf{x}}_0$  by $\tilde{\mathbf{x}}_0 = \mathcal{D}(\tilde{\mathbf{z}}_0)$. During sampling, the LDMs sample a noise latent vector $\mathbf{z}_T$ from Gaussian distribution \(\mathcal{N}(0, \mathbf{I})\). Subsequently, the trained LDM can utilize sampling methods like Denoising Diffusion Implicit Models (DDIM) \cite{song2020denoising} or DPM-Solver \cite{lu2022dpm} to obtain the latent representation of the sampled image $\mathbf{z}_0$ from $\mathbf{z}_T$ over $T$ timesteps. Then,  the decoder reconstructs the image from the latent by $\mathbf{x}_0=\mathcal{D}\left(\mathbf{z}_0\right)$. One can use methods like DDIM Inversion \cite{dimm_inversion} to invert the denoising process and recover the initial noise $\mathbf{z}_T$ from the generated image $\mathbf{x}_0$. 

\vspace{3pt}
\noindent
\textbf{Watermarks for Latent Diffusion Models.}
LDMs enable individuals to customize their models for specific styles of image generation via efficient training and fine-tuning using their data. Then, individuals can publish and exchange their models in the online market space, such as Civitai \cite{civitai} and Tensor.art \cite{tensorart}. However, these advancements have also raised concerns about the potential abuse of these models and the generated images. For example, the unauthorized distribution of these LDM models and their generated images leads to copyright infringement. Besides, it is difficult to trace the malicious users who generate realistic images to spread rumors and fake news on social media to potentially manipulating important social and economic events, such as political elections and the stock market. Therefore, enhancing LDMs with copyright protection and traceability techniques is crucial. Watermarking has a long history of alleviating these issues via labeling image content \cite{Ruanaidh_Dowling_Boland_1996}, which involves incorporating watermark information into the generated images. Then, one can trace the original information of these images by verifying the watermark, such as the model and user that produced these images.

Existing watermarking methods for LDMs can be categorized into post-processing and in-generation-process watermarks. Post-processing methods add watermarks to images after they have been generated by LDMs. For instance, the Stable Diffusion repository provides methods like DWT-DCT \cite{ rahman2013dwt} and RivaGAN~\cite{zhang2019robust}. Despite their widespread usage, direct modification to the images can degrade image quality \cite{fernandez2023stable}. Alternatively,  recent research proposes in-generation-process watermarks, which integrate the watermark embedding with the image generation process. Stable Signature~\cite{fernandez2023stable}  and AquaLora \cite{feng2024aqualora} embed watermarks by fine-tuning the VAE decoder and U-Net of the LDMs, respectively. These model-based methods improve the watermarked image quality but they require finetuning of the LDM, whcih can potentially degrade LDMs' performance.

Recently, researchers have explored latent-based watermarks, which embed the watermarks into the latent space of the diffusion models. Tree-Ring \cite{wen2023tree} encodes the watermark in the frequency domain of latent noise. Gaussian Shading \cite{yang2024gaussian} embeds watermarks into the latent variable while ensuring it follows a Gaussian distribution. DiffuseTrace \cite{lei2024diffusetrace} proposes to use an encoder model to generated watermarked latent variable. PRC-Watermark achieves undetectable watermarking by selecting the initial latent variables using a pseudorandom error-correcting code~\cite{christ2024pseudorandom}.  
Compared to model-based methods, latent-based methods are more user-friendly in practice because they avoid model retraining, maintain image quality, and offer plug-and-play usability. 

However, our research reveals a critical issue of these methods: most of them generate one watermark for one specific user; as a result, all the generated images from this user will have the same constant watermark signal even though it is invisible. This uniformity undermines the stealthiness of the watermarks, increasing the risk of copyright infringement.
Some methods, such as Gaussian Shading, has randomness by encrypting the watermarked latent noise with stream ciphers, but this approach requires a unique nonce for each watermarked latent noise, meaning every generated image must be paired with a nonce for watermark verification, which adds significant overhead in high-volume scenarios. 
PRC-Watermark also shows some stealthiness by selecting the initial latent using a pseudorandom error-correcting code but lacks traceability, limiting its applicability in multi-user environments. We address these issues and propose a more effective and stealthy watermark mechanism.

%-------------------------------------------------------------------------------
\section{Watermark Probe Attack}
\label{sec:watermark_detection_attack}

\begin{figure}[t]
    \centering
    \includegraphics[width=\linewidth]{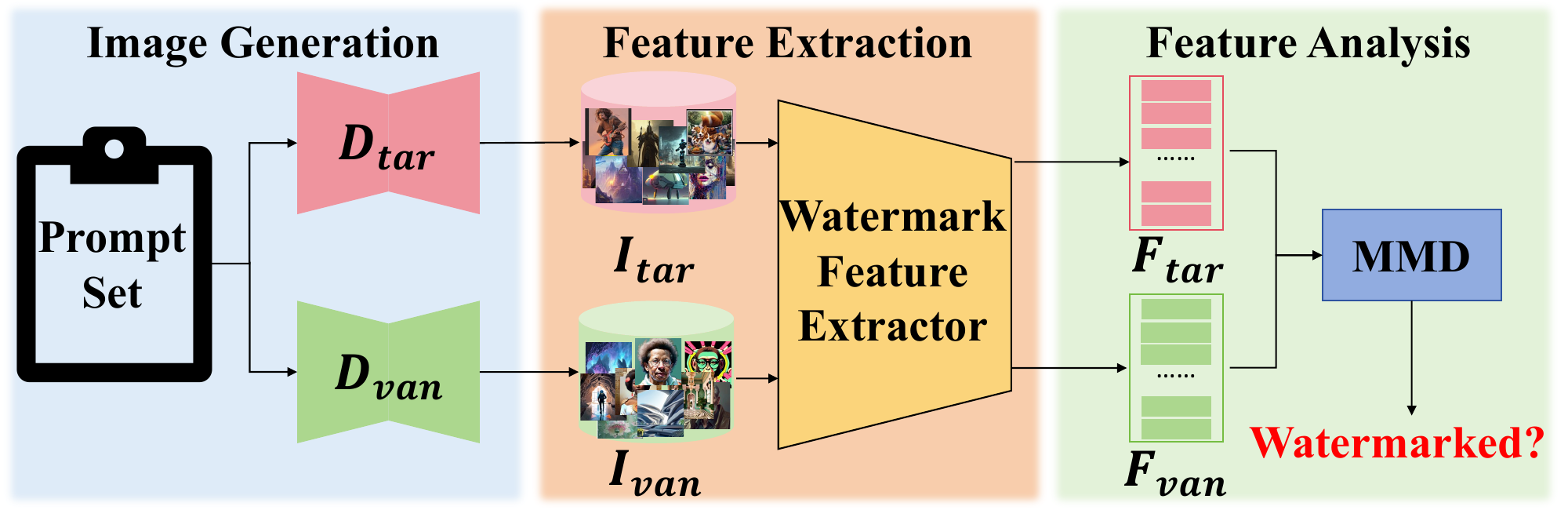}
    % \vspace{-0.6cm}
    \caption{The pipeline of our watermark probe attack.}
    \label{fig:pipeline}
    % \vspace{-0.6cm}
\end{figure}

We introduce a promising watermark probe attack to detect the existence of latent-based watermarks without the access to the corresponding secret watermark decoder. 

\subsection{Threat Model}
\label{sec:threat_model}
The watermark probe attack targets a scenario with two parties: the LDM owner that provides the image generation service and the attacker whose goal is to detect whether watermark exists in the generated images.

\vspace{3pt}
\noindent
\textbf{Model Owner.}
The LDM owner typically develop their models by training from a open-source LDM using their own dataset. Then, they can profit from their LDMs by sharing their models on online platforms such as Huggingface \cite{huggingface} , Civitai \cite{civitai} , and Tensor.Art \cite{tensorart}  or providing image generation services through API access. To prevent copyright infringement and ensure traceability in cases of misused images, the owner embeds imperceptible watermarks in each generated image without degrading image quality. With the secret watermark decoder,  the owner can verify whether the images are generated by their own models for copyright verification. Without the secret decoder, one cannot detect and recover the embedded watermark from images.

\vspace{3pt}
\noindent
\textbf{Attacker.} The attacker's goal is to redistribute the LDM-generated images and use these images to forge fake information without being traced for legal consequences. Therefore, they need to know whether the images contain a watermark. If no watermark's presence is detected, the attacker can use these images directly. If the watermark's existence is detected, they can either give up these images or apply some techniques to remove the watermark. 
The attacker's settings to detect the presence of watermarks from the generated images are the following: the attacker can generate images using model owner's API service and control only the prompts; the attacker does not know the watermarking method and the secret watermark decoder. Besides, the attacker has access to the open-source LDM, which approximates the watermark-free version of the target LDM. 

\subsection{Attack Pipeline}
\label{sec:wpa_overview}

As discussed in previous section, the latent-based watermark methods embed a constant signal in all the generated images for one specific watermark. Therefore, we attempt to extract this constant signal when the target LDM is watermarked. If the target LDM is watermark-free  (refer as vanilla in this paper ) , we cannot extract this constant signal. Since most customized LDMs are fine-tuned from the open-source LDMs with the model owner's data, we can use the open-source LDM to approximate the vanilla version of the target LDM. Based on the above analysis,  we propose the pipeline of our watermark probe attack as shown in Fig.~\ref{fig:pipeline}, which consists of three modules: Image Generation, Feature Extraction, and Feature Analysis. The Image Generation module generate the target image set \( I_{{tar}} \) and the vanilla image set
\( I_{van} \) from the target LDM and the vanilla LDM using the same prompts \( P \), respectively. These two image sets are used in the Feature Extraction module to train a Watermark Feature Extractor $WFE$ to output target feature \( F_{{tar}} \) and vanilla feature \( F_{{van}} \), respectively. We propose the designated loss functions so that $WFE$ converges only when the watermark is present in the target image set. After training, we will evaluate the distribution difference between \( F_{{tar}} \) and  \( F_{{van}} \)  to determine whether the target LDM is watermarked.

\vspace{3pt}
\noindent
\textbf{Image Generation.} This module aims to generate two image sets for detecting the constant watermark signal when the target LDM is watermarked. More specifically, we use a prompt set \( P \) to generate the target image set \( I_{{tar}} \) and the vanilla image set \( I_{{van}} \) from the target LDM \( D_{{tar}} \) and the vanilla LDM \( D_{{van}} \), respectively. Ideally, the watermark-free version of the target model would be used as \( D_{{van}} \)  to generate \( I_{{van}} \) such that the only difference between the two sets is the presence of the watermark. Since most LDMs are fine-tuned from open-source models \cite{10377881,10.1145/3658170}, there is a similarity between the output distributions of the target and vanilla LDMs.  Therefore, in practice, we can use the open-source LDM to approximate  \( D_{{van}} \) so that the differences caused by the watermark can be effectively detected in the next module.

\vspace{3pt}
\noindent 
\label{sec:feature_extraction}
\textbf{Feature Extraction.} In this module, we attempt to detect the watermark signal in the generated images by training a Watermark Feature Extractor $WFE$. When the watermark exists in the target images, the extractor should learn to extract the target feature \(F_{tar}\) with decreasing variance because it tries to identify the constant watermark signal; besides, the extractor should also identify the significant difference between the \(F_{van}\) and \(F_{tar}\) because the watermark is embeded in \(D_{tar}\)  instead of \(D_{van}\). On the contrary, if there is no watermark exists in the target images, \(F_{tar}\) should appear as a random distribution, and the distribution of \(F_{tar}\) and \(F_{van}\) should be similar. To encourage these behaviors, we first introduce \textbf{A}ggregating loss for \textbf{T}arge feature \(\mathcal{L}_{at}\), which calculates the variance of the target features as shown in Eq.~\ref{equ:Lat}:
\begin{equation}
\mathcal{L}_{at}=\frac{1}{N(N-1)} \sum_{i=1}^{N} \sum_{j=1, j \neq i}^{N} \left\|f_{{tar}}^i-f_{{tar}}^j\right\|^2.
\label{equ:Lat}
\end{equation}
\(\mathcal{L}_{at}\) encourages $WFE$ to extract the constant signal when watermark exists  in the target images  \(I_{tar}\). 
In addition, we introduce \(\mathcal{L}_{dtc}\) loss to \textbf{D}istinguish the difference between the \textbf{T}arget feature \(F_{tar}\) and \textbf{C}lean feature \(F_{van}\).  \(\mathcal{L}_{dtc}\)  calculates the reciprocal of the difference between the matched \(f_{tar}\) and \(f_{van}\) as shown in Eq.~\ref{equ:Ldtc}:
\begin{equation}
\mathcal{L}_{dtc} =\frac{1}{\frac{1}{N^2} \sum_{i=1}^{N} \sum_{j=1}^{N} \left\|f_{{tar}}^i-f_{{van}}^j\right\|^2} .
\label{equ:Ldtc}
\end{equation}
\(\mathcal{L}_{dtc}\) encourages $WFE$ to identify the difference between \(I_{tar}\) and \(I_{van}\)  when watermark exists  in the  \(I_{tar}\). Idealy, both \(\mathcal{L}_{at}\) and  \(\mathcal{L}_{dtc}\) will converge when the \(D_{tar}\) is watermarked and vice versa. However, since we use the open-source LDM to approximate \(D_{van}\),   it is possible for \(\mathcal{L}_{at}\) and  \(\mathcal{L}_{dtc}\) to converge when \(D_{tar}\) is watermark-free because $WFE$ treat the inherent model difference as the watermark difference. To alleviate this, we propose the \textbf{G}uassian \textbf{C}onstraint loss \(\mathcal{L}_{gc}\), motivated by one property of watermarking: when the input is a watermark-free image, the watermark extractor should produce a random output because no constant watermark signal exists in the images. Hence, \(\mathcal{L}_{gc}\) calculates the KL divergence \cite{KLDivergence} between the \(F_{van}\)'s distribution and uniform distribution, as shown in Eq.~\ref{equ:Lmc}.
\begin{equation}
\mathcal{L}_{gc} =\frac{1}{N} \sum_{i=1}^{N} \text{KL}\left(f_{{van}}^i \parallel \frac{1}{M} \right), \label{equ:Lmc}
\end{equation}
where \( N \) is the batch size, \( M \) is the feature dimension size. In summary, only when \(D_{tar}\) is watermarked these three types of losses will converge together, resulting significant distribution difference between   \(F_{tar}\) and  \(F_{van}\). 

\vspace{3pt}
\noindent
\textbf{Feature Analysis.} The last module determines whether the target LDM \( D_{tar} \) contains a watermark. We use the trained  $WFE$ from previous method to extract  \( F_{van} \) and \( F_{tar} \) from  \( I_{van} \) and  \( I_{tar} \). Then, we measure the distribution difference between these two features using metric like Maximum Mean Discrepancy (MMD) \cite{MMD} .  If the feature distributions of \( F_{van} \) and \( F_{tar} \)  exhibit significant differences, our approach predicts that the target model has been watermarked.

\begin{figure*}[t]
  \centering
  \includegraphics[width=.98\linewidth]{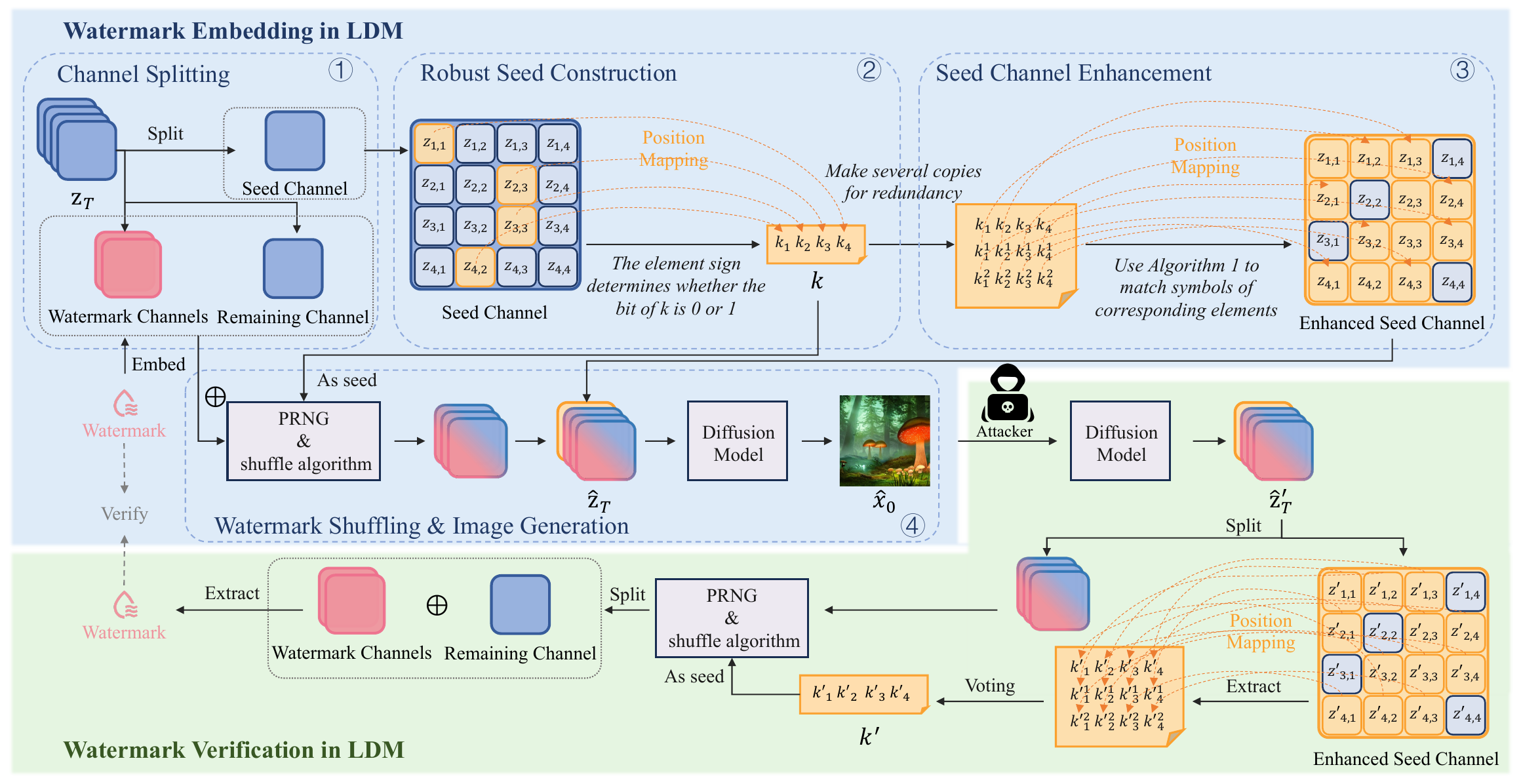}
  % \includegraphics[width=\linewidth]{Figure/SWA-LDM.pdf}
  % \vspace{-0.3cm}
  \caption{The framework of \tool. 
  We extract the seed $\mathbf{k}$ from randomly initialized latent variables and use it to shuffle the remaining latent variables where the watermark is inserted. This ensures the watermark information is randomized in each generated image. 
  % This is followed by a denoising process to generate the watermarked images $x_0$. For extraction, it is sufficient to apply diffusion inversion and reverse all the operations mentioned above.
  }
  \label{fig:swa-ldm}
  % \vspace{-0.5cm}
\end{figure*} 

%-------------------------------------------------------------------------------
\section{\tool{}}

Existing latent-based watermarking methods embed identical watermark patterns across distinct generated images, making them susceptible to watermark probe attacks. 
To address this vulnerability, 
we propose \tool, a lightweight framework that can be integrated with any existing latent-based watermarking method to enhance stealthiness. \tool introduces \textit{dynamic watermark randomization} and randomizes the watermark information embedded in each image by leveraging the inherent randomness of latent noise. This approach fundamentally disrupts pattern recognition in probe attacks by eliminating static watermark signatures. The details are in the following.

\subsection{Overview} 
The \tool framework operates through two phases (see Fig.~\ref{fig:swa-ldm}). The \textbf{watermark embedding phase} begins by first by sampling Gaussian latent noise \(\mathbf{z}_T\), embedding a watermark via existing methods (e.g., Gaussian Shading), then dynamically randomizing the embedded watermark using the latent noise's inherent variability (a separated seed channel) to produce the stealth-optimized latent \(\hat{\mathbf{z}}_T\). The image $\mathbf{x}_0$ is then synthesized via standard denoising, ensuring compatibility with unmodified LDMs. 
During the \textbf{watermark verification phase}, we use LDM's encoder and the diffusion inversion technique to reconstructs the stealth-optimized latent \(\hat{\mathbf{z}}'_T\), from which the seed channel is isolated to decode the watermark.  

\subsection{The Watermark Embedding Phase}
\label{sec:watermark_embedding}

To achieve dynamic watermark randomization and obtain the stealth-optimized latent \(\hat{\mathbf{z}}_T\), we implements two key innovations: 1) \textit{Latent-driven randomization}: A dedicated channel is partitioned from the Gaussian-distributed latent noise 
\({\mathbf{z}}_T\) to derive a randomization seed $\textbf{k}$. This seed dynamically shuffles the embedded watermark, eliminating fixed patterns. Unlike prior methods requiring external randomness storage, \tool exploits the inherent randomness already present in the latent space—a resource naturally recoverable during diffusion inversion, avoiding storage overhead.
2) \textit{Noise-resilient seed encoding}: 
As this seed is crucial for recovering of the original watermark during the verification process, we enhance its reliability with robust seed construction, seed channel enhancement, and robust seed extraction to counteract any potential noise caused by the image transmission or diffusion inversion process. 
These strategies ensures both stealth (via latent-driven randomization) and robustness (via noise-resistant encoding), while maintaining compatibility with existing latent-based watermarking pipelines.
We illustrate detailed steps as follows:

\vspace{3pt}
\noindent
\textbf{Channel Splitting.}
To achieve latent-driven randomization, we split the latent noise \( \mathbf{z}_T \) into three parts: seed channels \( \mathbf{z}_T^k \in \mathbb{R}^{c_k \times h \times w} \), watermark channels \( \mathbf{z}_T^w \in \mathbb{R}^{c_w \times h \times w} \), and the remaining channels \( \mathbf{z}_T^n \in \mathbb{R}^{c_n \times h \times w} \). The watermark channel is used for watermark embedding using a chosen latent-based watermarking method (\eg, \cite{yang2024gaussian, wen2023tree, lei2024diffusetrace}), while the seed channel is used for random seed construction. Note that the remaining channels are optional, depending on the number of channels the original watermarking method employs for embedding the watermark.

\vspace{3pt}
\noindent
\textbf{Robust Seed Construction.}
Due to transmission noises (e.g., perturbations like JPEG compression on image $x_0$) or inaccuracies in the diffusion inversion steps, the reconstructed \( \mathbf{z}_T' \) may not perfectly match the original \( \mathbf{z}_T \). This present challenges for us to reliably construct the seed \( \mathbf{k} \) from \( \mathbf{z}_T' \). 
To address this, \tool abstracts specific elements from the seed channel to construct each bit of \( \mathbf{k} \). 

First, we define a mapping function \( \mathcal{M} \) that consistently selects fixed locations in the seed channel for each bit in \( \mathbf{k} \). Specifically, \( \mathcal{M} \) maps each bit index to a unique position \((i, j, q)\) in \( \mathbf{z}_T^k \), where \( i \) represents the channel index, and \( (j, q) \) denote spatial coordinates. Given \( N = c_k \times h \times w \), this ensures that \tool always accesses the same positions in \( \mathbf{z}_T^k \) during \( \mathbf{k} \)-bit construction. For simplicity, we implement \( \mathcal{M} \) as a sequential mapping that unfolds \( \mathbf{z}_T^k \) linearly, ensuring a deterministic and repeatable key construction process.

Next, each bit of \( \mathbf{k} \) is sampled based on the sign of specific latent variables \( z^k_{T, i, j, q} \) within \( \mathbf{z}_T^k \). Letting \( M \) denote the bit-length of \( \mathbf{k} \), each bit is determined as follows: 
\begin{equation}
\mathbf{k} = \left[ k_1, \dots, k_{M} \right] \quad k_m = \begin{cases}
1, \text{if } z^k_{T, i_m, j_m, q_m} > 0 \\
0, \text{if } z^k_{T, i_m, j_m, q_m} \leq 0,
\end{cases}
\end{equation}
where \((i_m, j_m, q_m) = \mathcal{M}(m)\) indicates the index of \(k_m\) in the seed channel \(\mathbf{z}_T^k\).

\newcommand\mycommfont[1]{\footnotesize\ttfamily\textcolor{brown}{#1}}
\SetCommentSty{mycommfont}

\begin{algorithm}[t]
    \caption{Seed Channel Enhancement}
    \label{alg:generate_key}
    \SetAlgoNlRelativeSize{-1}
    \SetNoFillComment
    \renewcommand{\baselinestretch}{0.6} % Adjust line spacing
    % \scriptsize
    \footnotesize
    
    \KwIn{$\mathbf{z}_T^k$: Latent noise in seed channel, $\mathbf{k}$: Extracted seed bits, $R$: Number of redundancies, $\mathcal{M}$: Mapping function}
    \KwOut{$\overline{\mathbf{z}}_T^{k}$: Modified seed channel with robust seed information}
            
    % \SetKwFunction{FMain}{ExtractAndConstructKey}
    % \SetKwProg{Fn}{Function}{:}{}
    % \Fn{\FMain{$\mathbf{z}^k$, $M$, $M$}}{
        \For{$r \gets 1$ \KwTo $R$}{
            $M \gets \text{len}(\mathbf{k})$
            \For{$m \gets 1$ \KwTo $M$}{
                % \tcc{Find the latent noise corresponding to $k^r_m$}
                \textcolor{brown}{/*~Find the latent noise corresponding to $k^r_m$~*/}\\
                $(i, j, q) \gets \mathcal{M}(r \times M + m)$ \\
                $k^r_m$ $\gets 1$ \text{ if } ${z}^k_{T, i, j, q} > 0$ \text{ else } $0$\\
                
                \If{$k^r_m \neq k_m$}{
                    % \tcc{Search for latent noise to swap}
                    \textcolor{brown}{/*~Search for latent noise to swap~*/}\\
                    $p \gets m + 1$\\
                    \While{True}{
                        $(i', j', q') \gets \mathcal{M}(r \times M + p)$\\
                        {new\_bit} $\gets 1$ \text{ \textbf{if} } ${z}^k_{T, i', j', q'} > 0$ \text{ \textbf{else} } $0$\\
                        
                        \If{new\_bit $= k_m$}{
                            \text{swap}(${z}^k_{T, i, j, q}$, ${z}^k_{T, i', j', q'}$)\\
                            Break\\
                        }
                        $p \gets p + 1$\\
                    }
                }
            }
        }
        $\overline{\mathbf{z}}_T^{k} \gets \mathbf{z}_T^{k}$\\
        
        \Return{$\overline{\mathbf{z}}_T^{k}$}
    % }
\end{algorithm}

\vspace{3pt}
\noindent
\textbf{Seed Channel Enhancement.}
While the seed construction accounts for noise variations, it may still fail to reliably recover \( \mathbf{k} \) under perturbations. To address this, we propose a method to construct redundant seed information within \(\mathbf{z}_T^k\), ensuring robust seed extraction with minimal modification to \(\mathbf{z}_T^k\).
Let \( R \) denote the number of redundant seeds, with the \( r \)-th redundant seed represented as \( \mathbf{k}^r \) for \( r \in [1, R] \). Each bit of the redundant seed satisfies \( k^r_m = k_m \).
For each redundant seed \(\mathbf{k}^r\), we map it to a set of seed channel using the mapping function \(\mathcal{M}\). The latent variable \(z^k_{T, i_n, j_n, q_n}\) corresponds to \(k^r_{m}\) with \((i_n, j_n, q_n) = \mathcal{M}(r \times M + m)\). If the relationship between \(k^r_{m}\) (either 0 or 1) and \(z^k_{T, i_n, j_n, q_n}\) (either \(\leq 0\) or \(> 0\)) does not match, we search for a seed channel element that satisfies the condition and swap the corresponding values.
The seed channel enhancement process is detailed in Algorithm~\ref{alg:generate_key}. This algorithm takes the seed channel \(\mathbf{z}_T^k\), the seed \(\mathbf{k}\), and the number of redundant seed \(R\) as input, and outputs the enhanced seed channel \(\overline{\mathbf{z}}_T^{k}\), which includes the redundant seed. 

\vspace{3pt}
\noindent
\textbf{Watermark Shuffling.}
As previously discussed, \tool embeds the watermark into the watermark channel, resulting in \(\hat{\mathbf{z}}_T^w\). To ensure both randomization and recoverability of the watermark channel, \tool employs a pseudorandom number generator (PRNG) and a shuffle algorithm. The PRNG uses a random seed to generate an unpredictable sequence, which is crucial for randomizing watermark embedding and enabling reliable reconstruction during verification. \tool uses \( \mathbf{k} \) as a seed for the PRNG(PCG64)~\cite{oneill:pcg2014}. The Fisher-Yates shuffle algorithm~\cite{Fisher_Yates-AFP} is then applied to permute \( \operatorname{concat}(\hat{\mathbf{z}}_T^w, \mathbf{z}_T^n) \), dispersing watermark information across the latent space. Finally, we concatenate the enhanced seed channel \(\overline{\mathbf{z}}_T^{k}\) with the shuffled watermark channel to form the final stealth-optimized watermarked latent \(\hat{\mathbf{z}}_T\).

\vspace{3pt}
\noindent
\textbf{Image Generation.}
After constructing the stealth-optimized watermarked latent \(\hat{\mathbf{z}}_T\), the image generation process follows the standard procedure of the LDMs. Specifically, we utilize DDIM~\cite{song2020denoising} for denoising of \(\hat{\mathbf{z}}_T\). Once the denoised latent \(\hat{\mathbf{z}}_0\) is obtained, the watermarked image \(\hat{\mathbf{x}}_0\) is generated by applying the LDM decoder \(\mathcal{D}\): \(\hat{\mathbf{x}}_0 = \mathcal{D}(\hat{\mathbf{z}}_0)\).

\subsection{The Watermark Verification Phase}
\noindent
\textbf{Diffusion Inversion.}
For watermark verification, we use the LDM encoder \(\mathcal{E}\) to map the watermarked image \(\hat{\mathbf{x}}_0\) back to the latent space, obtaining \(\hat{\mathbf{z}}'_0 = \mathcal{E}(\hat{\mathbf{x}}_0)\). We then apply diffusion inversion over \(T\) timesteps, estimating the additive noise to recover 
\(\hat{\mathbf{z}}'_T \approx \hat{\mathbf{z}}_T\). Here, DDIM inversion~\cite{dimm_inversion} is used to approximate the original latent noise.

\vspace{3pt}
\noindent
\textbf{Robust Seed Extraction.}  
With \(\hat{\mathbf{z}}'_T\) obtained, we partition it to isolate the seed channel \(\overline{\mathbf{z}}_T^{k'}\) containing the redundant seed information and the shuffled channel. Using the fixed mapping function \(\mathcal{M}\), we extract the redundant seed information from predetermined positions in  \(\overline{\mathbf{z}}_T^{k'}\) to obtain both the seed \(\mathbf{k}'\) and its redundant bits \(\{\mathbf{k}^{r'} \mid r \in [1, R]\}\). 
Each bit \(k'_m\) of \(\mathbf{k}'\) is determined by a majority voting mechanism, wherein if more bits are zero than one among \(k'_m\) and \(\{ k^{r'}_m | r \in [1, R] \}\), \(k'_m\) is set to zero; otherwise, it is set to one.

\vspace{3pt}
\noindent
\textbf{Watermark Reshuffling and Verification.}  
After recovering the seed \(\mathbf{k}'\), we use it as the seed for a pseudorandom number generator (PCG64) and reapply the Fisher-Yates shuffle algorithm to re-shuffle the latent noise, excluding the seed channel. 
This reshuffled latent noise is then split to isolate \(\hat{\mathbf{z}}_T^{w'}\) and \(\mathbf{z}_T^{n'}\). 
Finally, based on the latent-based watermarking method employed, we extract and verify the watermark from \(\hat{\mathbf{z}}_T^{w'}\).

%-------------------------------------------------------------------------------
\section{Experiment}
All experiments are implemented using PyTorch 2.0.1 and the Diffusers 0.24.0 library, running on a single NVIDIA A800 GPU.

\subsection{Setup}
\label{sec:experiment_setup}
\noindent
\textbf{Latent Diffusion Models.} 
We employ three widely-used Stable Diffusion models as base models: Stable Diffusion v1-5 (SD v1-5), Stable Diffusion v2-1 (SD v2-1), and SD-XL 1.0-base (SDXL 1.0). Since these models are open-source, both attackers and model owners have access to them.
To simulate real-world scenarios, we assume model owners fine-tune these base models to create customized models and apply watermarking techniques to protect both the models and their generated images. To mimic this setting, we download 60 fine-tuned checkpoints from Hugging Face \cite{huggingface}, with 20 checkpoints tuned from each base model. A complete list of these checkpoints is in Sec.~\ref{sec:models_list}.
Compared to previous works, our study covers the largest model set (60 models, \vs Tree-ring~\cite{wen2023tree} with 1, Gaussian Shading~\cite{yang2024gaussian} with 3, and DiffuseTrace~\cite{lei2024diffusetrace} with 2).

\vspace{3pt}
\noindent
\textbf{Image Generation Details.}  
We generate 512×512-pixel images with prompts from the widely-used Stable-Diffusion-Prompts dataset~\cite{Gustavosta}. The latent noise dimensions are set to 4×64×64, and the classifier-free guidance scale is 7.5. We employ DDIM sampling ~\cite{song2020denoising} with 50 timesteps. 
Since the original prompts of generated images are often unknown in practice, we use an empty prompt for diffusion inversion~\cite{dimm_inversion}, while setting the  classifier-free guidance scale to 1 and perform 50 timesteps of DDIM inversion.

\vspace{3pt}
\noindent
\textbf{Baselines.} 
We evaluate three representative latent-based watermarking methods: Tree-ring~\cite{wen2023tree}, Gaussian Shading~\cite{yang2024gaussian}, and DiffuseTrace~\cite{lei2024diffusetrace}, which respectively represent frequency-domain, spatial-domain, and encoder-based watermark embedding approaches in the latent space.
For Gaussian Shading, we test both implementations, with and without the ChaCha20~\cite{bernstein2008chacha} secure stream cipher, which shuffles the watermark sequence. The detailed introduction and parameter settings of baseline:
\begin{itemize}
\item{For Tree-Ring \cite{wen2023tree}, }
it embeds a carefully constructed watermark pattern in the Fourier space of the initial latent noise. Following the original paper, we set the watermark pattern to multiple concentric rings, where each ring maintains a constant value drawn from a Gaussian distribution. This design ensures rotation invariance and resilience against various image transformations while minimally deviating from an isotropic Gaussian distribution. The radius of the watermark pattern is set to 16 to balance generation quality and verification performance. We embed the watermark into one latent channel and vary the constant values along the rings to generate distinct watermarks.

\item{For Gaussian Shading \cite{yang2024gaussian},} 
we adopt the parameters recommended in the original paper to balance watermark capacity and robustness. 
% Specifically, we use 1/8 of the latent height, 1/8 of the latent width, and one channel for watermarking. This watermark (dimensions 1×8×8) is inserted into three latent noise channels.
Specifically, the watermark size is $1/8$ of the latent height, $1/8$ of the latent width, and one channel. For generated images with resolution $3 \times 512 \times 512$, the corresponding latent noise dimensions are $4 \times 64 \times 64$, resulting in watermark dimensions of $1 \times 8 \times 8$. During embedding, To unify the setup with \tool, the watermark is redundantly replicated and inserted into three latent noise channels to enhance robustness.

\item{For DiffuseTrace \cite{lei2024diffusetrace}, }
% we per-train a DiffuseTrace Encoder-Decoder to generate a 3-channel latent noise (dimensions 3×64×64). We further fine-tune this Encoder-Decoder for each checkpoint to initialize watermarked latent noise tailored to each model. Decoder can extract 48 bit secret from latent noise
We use the publicly available code to obtain the DiffuseTrace Encoder-Decoder architecture, and pre-train the Encoder-Decoder to generate 3-channel latent noise (dimensions $3 \times 64 \times 64$) containing the DiffuseTrace watermark. To meet the input requirement of $4 \times 64 \times 64$ latent noise for Stable Diffusion (SD) models and unify the setup with \tool, we concatenate the 3-channel watermarked latent noise with a $1 \times 64 \times 64$ lantent noise sampled from a Gaussian distribution. To ensure compatibility with different SD models, we fine-tune the Encoder-Decoder for each specific SD model to initialize latent noise tailored to the model.  Following the original implementation, the bit length of the watermark is set to 48 during both training and testing.
\end{itemize}

\vspace{3pt}
\noindent
\textbf{Evaluation Metrics.} 
We evaluate watermarking methods based on three aspects: stealthiness, effectiveness, and image quality.
{For watermark stealthiness}, we evaluate the resilience of watermarks against watermark probe attacks. Attack performance is measured using the area under the ROC curve (AUC), where a higher AUC indicates higher detectability. Thus, we define stealthiness as $(1-\text{AUC of watermark probe attack})$.
{For watermark effectiveness}, we report AUC, true positive rate at 1\% false positive rate (TPR@1\%FPR), and bit accuracy for encoded watermark information.
{For watermarked image quality}, we use the CLIP score~\cite{radford2021learning} and the Fréchet Inception Distance (FID)\cite{NIPS2017_heuselgans}. The CLIP score computed using OpenCLIP-ViT/G\cite{Cherti_2023_CVPR}, measures the alignment between generated images and prompts. FID evaluates the feature similarity between generated and original images. To compute FID, we use the MS-COCO-2017 dataset~\cite{lin2014microsoft}, which provides paired original images and prompts. For each base model, we generate 5,000 images using the dataset prompts and compare them with the corresponding original images.

\vspace{3pt}  
\noindent  
\textbf{Setup of Watermark Probe Attack.}  
We evaluate the watermark probe attack under the assumption that the attacker uses different base models (SD v1-5, SD v2-1, SDXL 1.0). Unless otherwise specified, results are reported as the average attack performance across these models. The attacker generates 1,000 clean images using a base model and 1,000 target images using the target model to train the watermark feature extractor. The architecture and training details of the extractor are provided in the Sec.~\ref{sec:arc_wfe}.

\vspace{3pt}
\noindent
\textbf{Setup of \tool{}.} 
We integrate SWA-LDM with three baseline methods: with Tree-Ring, DiffuseTrace, and Gaussian Shading, denoted respectively by \tool{}(T-R), \tool{}(D-T), and \tool{}(G-S). Each method uses a seed channel count of 1 to construct an 8-bit seed with 64 redundant bits. The number of watermark channels is set to 1 for \tool{}(T-R) and 3 for both \tool{}(D-T) and \tool{}(G-S).

\vspace{3pt}
\noindent
\textbf{Image Perturbation Settings}
\label{sec:per_set}
% These attacks and their parameter ranges are as follows: 
In Sec.~\ref{sec:robustness_exp}, we evaluate the robustness of SWA-LDM against seven common image perturbations, which simulate potential attacks. The types of perturbations and their respective parameter ranges are detailed as follows:
\begin{itemize}
\item {JPEG Compression}, where the image is compressed using quality factors (QF) set to \{100, 90, 80, 70, 60, 50, 40, 30, 20, 10\}; 
\item {Random Crop}, which retains a randomly selected region covering \{80\%, 90\%\} of the original image area, discarding the rest; 
\item {Random Drop}, where randomly selected regions covering \{10\%, 20\%, 30\%, 40\%, 50\%\} of the image area are replaced with black pixels; 
\item {Resize and Restore} (Resize), where the image is resized to \{20\%, 30\%, 40\%, 50\%, 60\%, 70\%, 80\%, 90\%\} of its original dimensions and then restored to the original size; 
\item {Gaussian Blur} (GauBlur), applied with blur radii \( r \) set to \{1, 2, 3, 4\}; 
\item {Median Filter} (MedFilter), using kernel sizes \( k \) of \{1, 3, 5, 7, 9, 11\}; 
\item {Brightness Adjustment}, which modifies the image brightness using brightness factors \{0, 2, 4, 6\}.
\end{itemize}

\begin{table*}[t]
% \vspace{-0.3cm}
\centering
\caption{Comparison of \tool and baselines. The watermark effectiveness is evaluated with AUC, TPR@1\%FPR, and bit accuracy. The quality of the generated images is assessed using FID and CLIP scores. The stealthiness represents the failure rate of the proposed watermark probe attacks. Left to right are LDMs fine-tuned from SD v1-5/SD v2-1/SDXL 1.0.}
% \vspace{-0.3cm}
\label{tab:base_performance}
\resizebox{\textwidth}{!}{%
\begin{tabular}{@{}cccccccc@{}}
\toprule
\multirow{2}{*}{\textbf{Methods}} & \multirow{2}{*}{\textbf{Nonce}} & \multicolumn{6}{c}{\textbf{Metrics}} \\ \cmidrule(l){3-8} 
 &  & \textbf{AUC $\uparrow$} & \textbf{TPR@1\%FPR $\uparrow$} & \textbf{Bit Acc. $\uparrow$} & \textbf{FID $\downarrow$} & \textbf{CLIP-Score $\uparrow$} & \textbf{Stealthiness $\uparrow$} \\ \midrule
No watermark & \XSolidBrush & - & - & - & 29.77/27.01/75.83 & 0.32/0.29/0.30 & - \\ \midrule
Tree-Ring~\cite{wen2023tree}

& \XSolidBrush & 0.99/0.99/0.99 & 0.98/0.99/0.99 & - & 30.53/28.32/78.97 & 0.32/0.29/0.30 & 0.20/0.21/0.22 \\
DiffuseTrace~\cite{lei2024diffusetrace}
& \XSolidBrush & 0.99/0.98/0.84 & 0.98/0.94/0.43 & 0.97/0.95/0.69 & 30.15/26.83/83.68 & 0.32/0.29/0.30 & 0.20/0.21/0.29 \\
Gaussian Shading~\cite{yang2024gaussian} 
& \XSolidBrush & 1.00/1.00/1.00 & 1.00/1.00/1.00 & 0.99/0.99/0.99 & 31.58/29.82/70.39 & 0.32/0.29/0.30 & 0.00/0.01/0.08 \\
$\text{G-S}_{ChaCha20}$~\cite{yang2024gaussian}
& \Checkmark & 1.00/1.00/1.00 & 1.00/1.00/1.00 & 0.99/0.99/0.99 & 29.69/27.21/75.83 & 0.32/0.29/0.30 & 0.42/0.50/0.47 \\ \midrule
\tool(T-R) & \XSolidBrush & 0.99/0.99/0.99 & 0.99/0.99/0.99 & - & 30.24/27.43/70.21 & 0.32/0.29/0.30 & 0.47/0.49/0.47 \\
\tool(D-T) & \XSolidBrush & 0.99/0.97/0.81 & 0.98/0.94/0.35 & 0.97/0.94/0.66 & 29.80/26.90/76.89 & 0.32/0.29/0.30 & 0.49/0.49/0.50 \\
\tool(G-S) & \XSolidBrush & 0.99/0.99/0.99 & 0.99/0.99/0.99 & 0.99/0.99/0.99 & 30.53/27.28/75.29 & 0.32/0.29/0.30 & 0.46/0.51/0.50 \\ \bottomrule

\end{tabular}
}
% \vspace{-0.3cm}
\end{table*}

\subsection{Comparison to Baseline Methods}
\label{sec:comparison_baselines}
We evaluate \tool in terms of stealthiness, effectiveness, and robustness. Experimental results show that \tool significantly improves watermark stealthiness while maintaining similar watermark effectiveness and robustness. 

\vspace{3pt}
\noindent
\textbf{Stealthiness.}
We conduct watermark probe attack experiments against \tool and baselines. The attack performance, summarized in the "Stealthiness" column of Tab.~\ref{tab:base_performance}, shows the average stealthiness achieved by each method against an attacker using different base models. 

The results show that watermark probe attacks effectively detect watermarks in baseline methods. \tool improves stealthiness against these attacks. Among baseline methods, Gaussian Shading has the lowest stealthiness, while DiffuseTrace and Tree-Ring offer slight improvements but remain vulnerable. Gaussian Shading with ChaCha20 increases stealthiness but requires costly per-image nonce management. In contrast, \tool achieves ChaCha20-level stealthiness without nonce dependency, integrating smoothly with DiffuseTrace, Tree-Ring, and Gaussian Shading. Further analysis on the base model’s impact on detection is in Sec.~\ref{sec:ablation_studies}.

\vspace{3pt}
\noindent
\textbf{Effectiveness.}
As detailed in Sec.~\ref{sec:experiment_setup}, 
we verified the effectiveness of \tool{} at 60 checkpoints.
each generating 1,000 images, resulting in 60,000 watermarked and 60,000 clean images per method. 
As shown in Tab.~\ref{tab:base_performance}, \tool maintains comparable watermark effectiveness (AUC, TPR@1\%FPR, and bit accuracy) compared to the baselines, with occasionally slight decrease which is considered acceptable due to the large benefit brought by the enhanced stealthiness. 
\tool also maintains the FID and CLIP scores, preserving LDM-generated image quality.

\begin{table*}
    % \vspace{-0.5cm}
    \centering
    \caption{Watermark Verification AUC under each image perturbation. Cr. \& Dr. refers to random crop and random drop.}
    \label{table:robustness}
    % \vspace{-.3cm}
    % \resizebox{.9\textwidth}{!}{%  
    \begin{tabular}{@{}cccccccc@{}}
    \toprule
    \textbf{Methods} & \textbf{JPEG} & \textbf{Cr. \& Dr.} & \textbf{Resize} & \textbf{GauBlur} & \textbf{MedFilter} & \textbf{Brightness} & \textbf{Avg} \\ \midrule

    Tree-Ring & 0.98 & 0.99 & 0.99 & 0.98 & 0.98 & 0.99 & 0.99 \\
    DiffuseTrace & 0.96 & 0.99 & 0.98 & 0.96 & 0.96 & 0.92 & 0.96 \\
    Gaussian Shading & 0.99 & 1.00 & 1.00 & 1.00 & 1.00 & 0.99 & 0.99 \\
    $\text{G-S}_{ChaCha20}$ & 0.99 & 1.00 & 1.00 & 1.00 & 1.00 & 0.99 & 0.99 \\ \midrule
    \tool(T-R) & 0.95 & 0.95 & 0.98 & 0.95 & 0.96 & 0.94 & 0.95 \\
    \tool(D-T) & 0.93 & 0.97 & 0.97 & 0.93 & 0.95 & 0.96 & 0.95 \\
    \tool(G-S) & 0.96 & 0.98 & 0.98 & 0.97 & 0.97 & 0.93 & 0.97 \\ \bottomrule
    
    \end{tabular}
    % }
% \vspace{-.25cm}
    % \vspace{-0.5cm}
\end{table*}

\vspace{3pt}
\noindent
\textbf{Robustness.}
\label{sec:robustness_exp}
To evaluate the robustness of \tool, we assess its performance under seven common image perturbations: JPEG compression, random crop, random drop, resize and restore (Resize), Gaussian blur (GauBlur), median filter (MedFilter), and brightness adjustments. The detailed perturbation strength are shown in the Sec.~\ref{sec:per_set}.
For each parameter setting of every perturbation, we used 2,000 images generated by the SD v1-5 to evaluate performance. The average verification AUC for each perturbation is reported in Tab~\ref{table:robustness}.
Results indicate that \tool has slight impact on the robustness. This occurs because \tool requires complete recovery of each bit in the seed to retrieve the watermark. Nevertheless, unless the image undergoes quality-compromising levels of perturbation, our watermark remains practical.

\subsection{Ablation Studies}
\label{sec:ablation_studies}

\begin{table}[t]
% \vspace{-0.1cm}
\centering
\caption{Impact of different clean SD models on the watermark probe attacks (Stealthiness $\uparrow$). Left to right are target LDMs fine-tuned from SD v1-5/SD v2-1/SDXL 1.0.}
% \vspace{-0.3cm}
\centering
\label{tab:ab_sd_version}
\resizebox{\columnwidth}{!}{%
\begin{tabular}{@{}cccc@{}}
\toprule
\multirow{2}{*}{\textbf{Methods}} & \multicolumn{3}{c}{\textbf{SD version used to generate the clean images}} \\ \cmidrule(l){2-4} 
 & SD v1-5 & SD v2-1 & SD-XL v1.0 \\ \midrule

Tree-ring & 0.24/0.25/0.27 & 0.16/0.15/0.25 & 0.22/0.22/0.15 \\
DiffuseTrace & 0.18/0.22/0.30 & 0.20/0.21/0.30 & 0.22/0.21/0.28 \\
Gaussian Shading & 0.01/0.01/0.10 & 0.00/0.03/0.08 & 0.00/0.01/0.06 \\
$\text{G-S}_{ChaCha20}$ & 0.44/0.54/0.50 & 0.38/0.40/0.43 & 0.45/0.57/0.50 \\ \midrule

\tool (T-R) & 0.48/0.51/0.48 & 0.47/0.49/0.46 & 0.46/0.47/0.47 \\
\tool (D-T) & 0.47/0.52/0.52 & 0.49/0.48/0.46 & 0.52/0.47/0.53 \\
\tool (G-S) & 0.43/0.47/0.52 & 0.50/0.51/0.49 & 0.46/0.54/0.50 \\ \bottomrule

\end{tabular}%
}
% \vspace{-0.3cm}
\end{table}

% \vspace{3pt}
\noindent
\textbf{Impact of the clean SD model on watermark probe attack.} 
We evaluated whether the effectiveness of the watermark probe attack is influenced by the base model used by the attacker to generate clean images. Results shown in Tab.~\ref{tab:ab_sd_version} indicate that the choice of base model has minimal impact on attack performance, demonstrating that the watermark detection attack remains effective without requiring knowledge related to the target model.

\begin{table}[t]
% \vspace{-0.5cm}
\caption{Impact of image quantities on the watermark probe attacks(Stealthiness $\uparrow$). Left to right are target LDMs fine-tuned from SD v1-5/SD v2-1/SDXL 1.0.
% The target checkpoints are fine-tuned from SD v1-5/SD v2-1/SDXL 1.0.
}
% \vspace{-0.3cm}
\label{tab:ab_clean_img_num}
\resizebox{\columnwidth}{!}{%
\begin{tabular}{@{}ccccc@{}}
\toprule
\multirow{2}{*}{\textbf{Methods}} & \multicolumn{4}{c}{\textbf{Image Quantity}} \\ \cmidrule(l){2-5} 
 & 500 & 1,000 & 1,500 & 2,000 \\ \midrule

Tree-ring & 0.25/0.33/0.19 & 0.20/0.21/0.22 & 0.21/0.17/0.21 & 0.22/0.17/0.18 \\
DiffuseTrace & 0.22/0.20/0.32 & 0.20/0.21/0.29 & 0.24/0.20/0.28 & 0.22/0.21/0.30 \\
Gaussian Shading & 0.03/0.01/0.08 & 0.00/0.01/0.08 & 0.01/0.00/0.07 & 0.04/0.00/0.07 \\
$\text{G-S}_{ChaCha20}$ & 0.42/0.46/0.43 & 0.42/0.50/0.47 & 0.43/0.48/0.53 & 0.42/0.47/0.54 \\ \midrule
\tool (T-R) & 0.44/0.45/0.52 & 0.47/0.49/0.47 & 0.41/0.48/0.51 & 0.50/0.48/0.45 \\
\tool (D-T) & 0.49/0.53/0.47 & 0.49/0.49/0.50 & 0.52/0.47/0.50 & 0.49/0.51/0.47 \\
\tool (G-S) & 0.42/0.48/0.52 & 0.46/0.51/0.50 & 0.41/0.48/0.52 & 0.50/0.45/0.52 \\ \bottomrule

\end{tabular}%
}
% \vspace{-0.2cm}
\end{table}

\vspace{3pt}
\noindent
\textbf{Impact of image quantity on watermark probe attack.} 
Following the setup in Sec.~\ref{sec:experiment_setup}, we varied the number of images generated by the watermark probe attacker to assess its effect on attack performance. Results shown in Tab~\ref{tab:ab_clean_img_num}, indicate that within our sampled range, the watermark probe attack's effectiveness remains stable regardless of image quantity. 

\begin{figure}[t]
    \centering
    \begin{minipage}{\linewidth}
        \centering
        \includegraphics[width=0.8\linewidth]{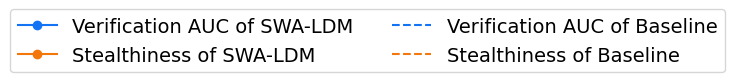} % 调整图例图片的宽度
        % \vspace{-0.3cm} 
    \end{minipage}
    
    \resizebox{\columnwidth}{!}{%
    \subfloat[\tool{}(G-S).]{\label{fig:redundancy_G-S}\includegraphics[width=0.35\linewidth]{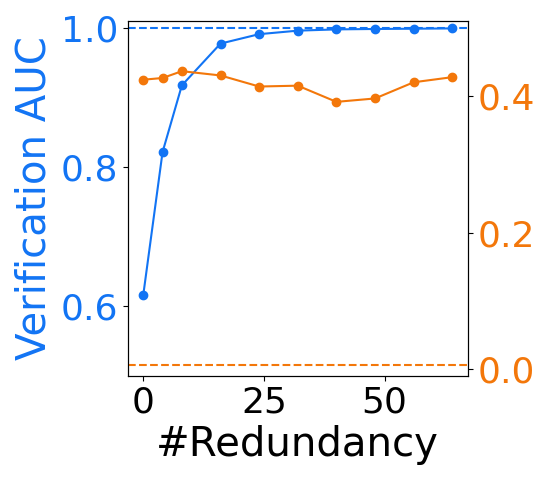}}\hspace{-0.015\linewidth}
     \subfloat[\tool{}(T-R).]{\label{fig:redundancy_T-R}\includegraphics[width=0.35\linewidth]{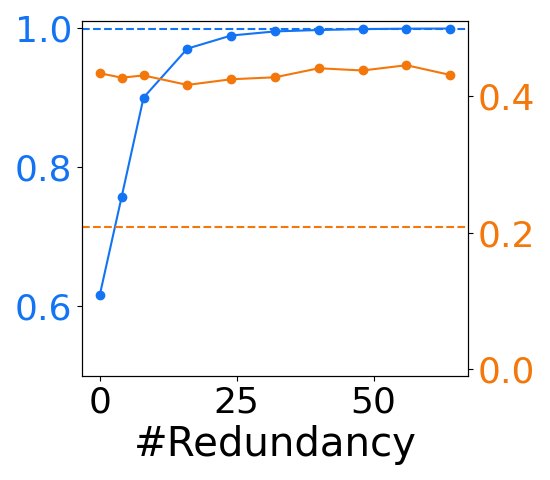}}\hspace{-0.015\linewidth}
     \subfloat[\tool{}(D-T).]{\label{fig:redundancy_D-T}\includegraphics[width=0.35\linewidth]{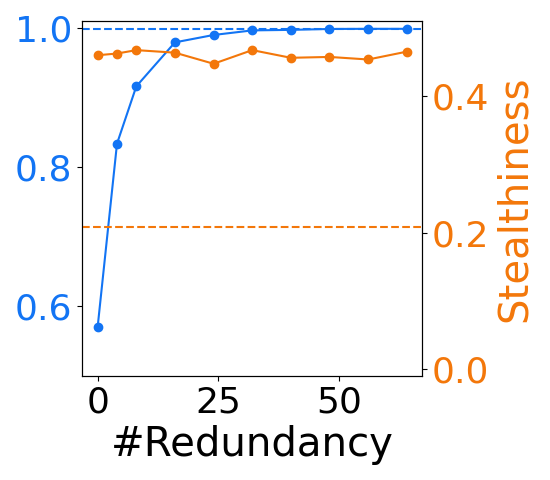}}
     }
     % \vspace{-0.3cm}
    \caption{Performance of \tool with varying numbers of redundancies. The effectiveness is demonstrated through AUC and stealthiness metrics, where (a) compares \tool(G-S) with Gaussian Shading and (b) compares \tool(T-R) with Tree-Ring. (c) compares \tool(D-T) with DiffuseTrace.}
    \label{fig:redundancy}
    % \vspace{-0.3cm}
\end{figure}

\vspace{3pt}
\noindent
\textbf{Impact of seed redundancy on stealthiness and verification performance.} 
% Using the experimental setup from Section~\ref{sec:e_detection}, we tested the influence of key redundancy on both watermark stealthiness and verification AUC. Show in \cref{fig:redundancy}.
Following the setup in Section~\ref{sec:experiment_setup}, we evaluate how varying seed redundancy levels affect watermark stealthiness and verification AUC. Results in Fig.~\ref{fig:redundancy} show that with minimal redundancy (4 redundancies), \tool achieves a verification AUC around 0.8, compared to a near-perfect verification AUC of 1 for watermarking methods without \tool, indicating an 80\% seed recovery success rate. As redundancy increases to 8, the recovery probability improves to 90\%, and with redundancy over 40, \tool achieves near-complete seed recovery without compromising verification AUC.  Across all redundancy levels, \tool maintains consistently high stealthiness.

\begin{figure}[t]
    \centering
    \begin{minipage}{\linewidth}
        \centering
        \includegraphics[width=0.8\linewidth]{Figure/ablation_studies/legend_image.png} % 调整图例图片的宽度
    \end{minipage}
    % \vspace{-0.1cm} 
    \resizebox{\columnwidth}{!}{%
    \subfloat[\tool{}(G-S).]{\label{fig:keybit_G-S}\includegraphics[width=0.35\linewidth]{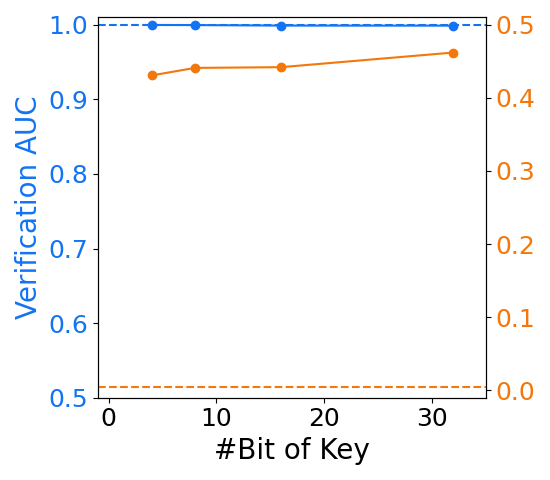}}\hspace{-0.015\linewidth}
     \subfloat[\tool{}(T-R).]{\label{fig:keybit_T-R}\includegraphics[width=0.35\linewidth]{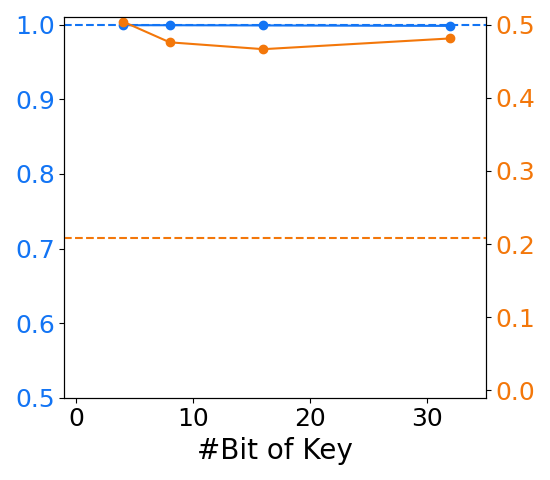}}\hspace{-0.015\linewidth}
     \subfloat[\tool{}(D-T).]{\label{fig:keybit_D-T}\includegraphics[width=0.35\linewidth]{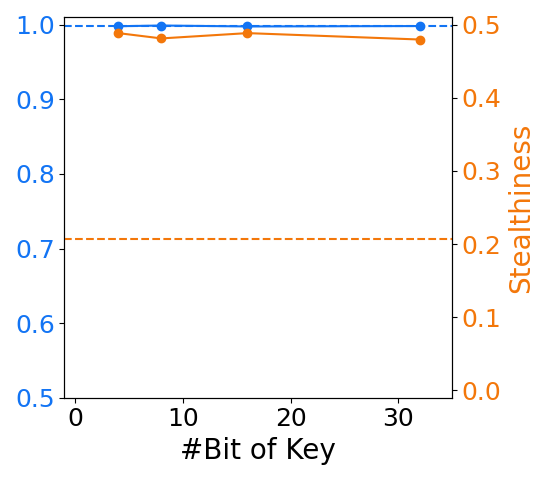}}
     }
     % \vspace{-0.3cm}
    \caption{Performance of \tool with varying bit number of seed. The effectiveness is demonstrated through AUC and stealthiness metrics, where (a) compares \tool(G-S) with Gaussian Shading and (b) compares \tool(T-R) with Tree-Ring. (c) compares \tool(D-T) with DiffuseTrace.}
    
    \label{fig:keybit}
    % \vspace{-0.6cm}
\end{figure}

\vspace{3pt}
\noindent
\textbf{Impact of bit number of seed on stealthiness and verification performance.} 
% Using the experimental setup from Section~\ref{sec:e_detection}, we tested the influence of key redundancy on both watermark stealthiness and verification AUC. Show in \cref{fig:redundancy}.
% Following the setup in Section~\ref{sec:experiment_setup}, we evaluate how varying bit number of key affects watermark stealthiness and verification AUC. Results in \cref{fig:redundancy} show that \tool maintains consistently high stealthiness.
In Sec.~\ref{sec:watermark_embedding}, we have introduced how \tool{} employs a pseudorandom number generator (PRNG) and a shuffle algorithm to randomize the watermarked latent noise. The seed is derived from the latent noise and serves as the seed for the PRNG. 
To analyze the impact of seed bit number on watermark stealthiness and verification performance, we have further evaluated \tool across a range of seed lengths from 4 to 32 bits, based on the setup described in Sec.~\ref{sec:experiment_setup}. 
Results in Fig.~\ref{fig:keybit}, indicate that both watermark stealthiness and verification AUC remain consistent regardless of the seed's bit number within this range. These findings suggest that the choice of seed length does not compromise the effectiveness or concealment of the watermark, providing flexibility in the design of the seed construction process.

\begin{table}[t]
    \centering
    \caption{Impact of seed construction on watermark verification AUC. The table compares results with (\Checkmark) and without (\XSolidBrush) the proposed seed construction mechanism. Left to right are LDMs fine-tuned from SD v1-5/SD v2-1/SDXL 1.0.}

    \label{tab:key_construction}
    % \vspace{-0.3cm}
    \resizebox{\columnwidth}{!}{%
    \begin{tabular}{@{}cccc@{}}
    \toprule
    \multirow{2}{*}{\textbf{\begin{tabular}[c]{@{}c@{}}Seed \\ Construction\end{tabular}}} & \multicolumn{3}{c}{\textbf{Watermark Methods}} \\ \cmidrule(l){2-4} 
    & \tool (T-R) & \tool (D-T) & \tool (G-S) \\ \midrule
    \XSolidBrush & 0.51/0.53/0.52 & 0.48/0.49/0.50 & 0.49/0.50/0.49 \\
    \Checkmark & 0.99/0.99/0.99 & 0.99/0.97/0.81 & 0.99/0.99/0.99 \\ \bottomrule
    
    \end{tabular}
    }
  % \vspace{-0.3cm}
\end{table}

\vspace{3pt}
\noindent
\textbf{Impact of seed construction.}
% As described in \cref{sec:watermark_embedding}, the key is constructed by sampling each bit based on the sign of specific latent variables. To evaluate the importance of this key construction mechanism, we have conducted an ablation study where the key construction process was replaced by directly using fixed latent variables as the key. 
As described in Sec.~\ref{sec:watermark_embedding}, the seed is constructed by sampling each bit from the sign of specific latent variables. To assess its importance, we have replaced this mechanism with fixed latent variables as the seed.  
% Following the experimental setup in Section~\ref{sec:experiment_setup}, the results are presented in \cref{tab:key_construction}. Evidently, the absence of the key construction mechanism significantly degrades performance. Without that, discrepancies between the estimated latent noise during diffusion inversion and the original latent noise prevented \tool{} from reconstructing the key correctly. This failure in key reconstruction rendered watermark verification infeasible, demonstrating the critical role of key construction in ensuring robust watermark performance.
Following the experimental setup in Sec.~\ref{sec:experiment_setup}, the results in Tab.~\ref{tab:key_construction} show that removing the seed construction significantly degrades performance. 
As analyzed in Sec.~\ref{sec:watermark_embedding}, during diffusion inversion, the reconstructed latent noise may not perfectly match the original latent noise, especially when the image experiences perturbations. These mismatches prevent accurate seed reconstruction, making watermark verification infeasible. This underscores the critical role of seed construction in maintaining robust watermarking.

\begin{table}[t]
    % \vspace{-0.3cm}
    \caption{Verification AUC with different sampling methods, including DDIM \cite{song2020denoising}, UniPC \cite{zhao2023unipc}, PNDM \cite{liu2022pseudo}, DEIS \cite{zhang2023fastsamplingdiffusionmodels}, and DPMSolver \cite{lu2022dpm}.}
    \label{tab:sample_method}
    % \vspace{-0.3cm}
    \resizebox{\columnwidth}{!}{%
    \begin{tabular}{@{}cccccc@{}}
    \toprule
    \multirow{2}{*}{\textbf{\begin{tabular}[c]{@{}c@{}}Watermark \\ Methods\end{tabular}}} & \multicolumn{5}{c}{\textbf{Sampling Methods}} \\ \cmidrule(l){2-6} 
     & DDIM  & UniPC  & PNDM  & DEIS  & DPMSolver \\ \midrule
    \tool (T-R) & 1.00 & 0.97 & 1.00 & 1.00 & 1.00 \\
    \tool (D-T) & 0.99 & 0.99 & 0.99 & 1.00 & 0.99 \\
    \tool (G-S) & 1.00 & 1.00 & 1.00 & 1.00 & 1.00 \\ \bottomrule
    
    \end{tabular}
    }
  % \vspace{-0.3cm}
\end{table}

\vspace{3pt}
\noindent
\textbf{Impact of sampling methods.}
% To validate generalization, we selected five commonly used sampling methods, As shown in \cref{tab:sample_method}, 
We tested five commonly used sampling methods. As shown in Tab.~\ref{tab:sample_method}, our method demonstrates stable watermark verification AUC across different sampling methods.

\begin{table}[t]
    \centering
    \caption{Verification AUC of \tool(T-R)/\tool(D-T)/\tool(G-S) with different denoising and inversion step.}
    % \vspace{-0.3cm}
    \resizebox{\linewidth}{!}{
    \begin{tabular}{@{}ccccc@{}}
    \toprule
    \multirow{2}{*}{\textbf{\begin{tabular}[c]{@{}c@{}}Denoising\\ Step\end{tabular}}} & \multicolumn{4}{c}{\textbf{Inversion Step}} \\ \cmidrule{2-5} 
     & 10 & 25 & 50 & 100 \\ \midrule

    10 & 0.99/0.99/1.00 & 1.00/1.00/1.00 & 1.00/1.00/0.99 & 1.00/1.00/0.99 \\
    25 & 1.00/0.99/1.00 & 1.00/0.99/1.00 & 1.00/1.00/1.00 & 1.00/1.00/1.00 \\
    50 & 1.00/1.00/1.00 & 1.00/1.00/1.00 & 1.00/1.00/1.00 & 1.00/1.00/1.00 \\
    100 & 1.00/1.00/1.00 & 1.00/1.00/1.00 & 1.00/1.00/1.00 & 1.00/1.00/1.00 \\  \bottomrule

    \end{tabular}
    }
    \label{tab:step}
    % \vspace{-0.7cm}
\end{table}
\vspace{3pt}
\noindent
\textbf{Impact of inversion step.}
In practice, the specific denoising step used in generation is often unknown, which can result in a mismatch with the inversion step. However, as shown in Tab.~\ref{tab:step}, this step mismatch does not affect the performance of our watermarking approach.

\section{Conclusion}

In conclusion, we expose the stealthiness issues in existing latent-based watermarking methods for Latent Diffusion Models (LDMs). We introduce a novel watermark probe attack that operates solely on generated images, setting a new standard in the field and highlighting the urgent need for enhanced watermarking strategies.
To counter these vulnerabilities, we present \tool, a plug-and-play component that enables the creation of stealthy, image-adaptive watermarks without incurring additional management costs. 
% Our approach leverages the randomness of latent noise to ensure robustness and uniqueness for each watermark. 
Comprehensive experiments validate the effectiveness of \tool in improving watermark stealthiness without compromising other watermark metrics. 

% \section*{Acknowledgments}
% This should be a simple paragraph before the References to thank those individuals and institutions who have supported your work on this article.

% {\appendix[Proof of the Zonklar Equations]
% Use $\backslash${\tt{appendix}} if you have a single appendix:
% Do not use $\backslash${\tt{section}} anymore after $\backslash${\tt{appendix}}, only $\backslash${\tt{section*}}.
% If you have multiple appendixes use $\backslash${\tt{appendices}} then use $\backslash${\tt{section}} to start each appendix.
% You must declare a $\backslash${\tt{section}} before using any $\backslash${\tt{subsection}} or using $\backslash${\tt{label}} ($\backslash${\tt{appendices}} by itself
%  starts a section numbered zero.)}

{
\appendices
\section{Evaluated Models.}
\label{sec:models_list}
% We have employed three widely-used Stable Diffusion models as base models: Stable Diffusion v1-5 (SD v1-5), Stable Diffusion v2-1 (SD v2-1), and SD-XL 1.0-base (SDXL 1.0). 
% For customized models, we have downloaded 60 checkpoints from Hugging Face \cite{huggingface}, fine-tuned from three base models (SD v1-5, SD v2-1, and SDXL 1.0). Detailed information on these 60 checkpoints is provided in \cref{tab:name_model}.
We utilize three widely-used Stable Diffusion models as base models: Stable Diffusion v1-5 (SD v1-5), Stable Diffusion v2-1 (SD v2-1), and SDXL 1.0-base (SDXL 1.0). Additionally, we download 60 checkpoints from Hugging Face \cite{huggingface}, which include models fine-tuned from these base models or equipped with adapters. These customized models, comprising either fine-tuned versions or base models enhanced with adapters, were used in our experiments. A list of the adapters and fine-tuned models can be found in Tab.~\ref{tab:name_model}.

\begin{table*}[ht]
% \vspace{-0.3cm}
\caption{The names of the 60 checkpoints used in our experiment}
\resizebox{\textwidth}{!}{%
\begin{tabular}{@{}clll@{}}
\toprule
\textbf{Type} & \textbf{base on runwayml/stable-diffusion-v1-5}                       & \textbf{base on stabilityai/stable-diffusion-2-1}    & \textbf{base on stabilityai/stable-diffusion-xl-base-1.0} \\ \midrule
\multirow{10}{*}{Adapters}  & latent-consistency/lcm-lora-sdv1-5                                    & sahibnanda/anime-night-vis-sd                        & alvdansen/BandW-Manga                                     \\
                            & Melonie/text\_to\_image\_finetuned                                    & sahibnanda/anime-real-vis-night                      & nerijs/pixel-art-xl                                       \\
                            & Kvikontent/midjourney-v6                                              & dlcvproj/cartoon\_sd\_lora                           & latent-consistency/lcm-lora-sdxl                          \\
                            & h1t/TCD-SD15-LoRA                                                     & jainr3/sd-diffusiondb-pixelart-v2-model-lora         & alvdansen/littletinies                                    \\
                            & ostris/depth-of-field-slider-lora                                     & dlcvproj/retro\_sd\_lora                             & Pclanglais/Mickey-1928                                    \\
                            & Norod78/sd15-megaphone-lora                                           & lora-library/lora-dreambooth-sample-dog              & artificialguybr/ColoringBookRedmond-V2                    \\
                            & rocifier/painterly                                                    & lora-library/artdecodsgn                             & fofr/sdxl-emoji                                           \\
                            & artificialguybr/pixelartredmond-1-5v-pixel-art-loras-for-sd-1-5       & nakkati/output\_dreambooth\_model\_preservation      & alimama-creative/slam-lora-sdxl                           \\
                            & patrickvonplaten/lora\_dreambooth\_dog\_example                       & Mousewrites/charturnerhn                             & Adrenex/chamana                                           \\
                            & artificialguybr/stickers-redmond-1-5-version-stickers-lora-for-sd-1-5 & lora-library/alf                                     & alvdansen/midsommarcartoon                                \\ \midrule
\multirow{10}{*}{Finetunes} & mhdang/dpo-sd1.5-text2image-v1                                        & ptx0/pseudo-flex-v2                                  & mhdang/dpo-sdxl-text2image-v1                             \\
                            & iamkaikai/amazing-logos-v2                                            & Vishnou/sd-laion-art                                 & Bakanayatsu/Pony-Diffusion-V6-XL-for-Anime                \\
                            & stablediffusionapi/counterfeit-v30                                    & n6ai/graphic-art                                     & Bakanayatsu/ponyDiffusion-V6-XL-Turbo-DPO                 \\
                            & Bakanayatsu/cuteyukimix-Adorable-kemiaomiao                           & artificialguybr/freedom                              & Lykon/dreamshaper-xl-lightning                            \\
                            & iamanaiart/meinamix\_meinaV11                                         & bguisard/stable-diffusion-nano-2-1                   & Lykon/dreamshaper-xl-v2-turbo                             \\
                            & stablediffusionapi/maturemalemix-v14                                  & cloudwithraj/dogbooth                                & Lykon/AAM\_XL\_AnimeMix                                   \\
                            & Lykon/DreamShaper                                                     & bghira/pseudo-flex-v2                                & Linaqruf/animagine-xl-2.0                                 \\
                            & Lykon/AnyLoRA                                                         & WildPress/simba\_model                               & fluently/Fluently-XL-v4                                   \\
                            & simbolo-ai/bagan                                                      & nishant-glance/model-sd-2-1-priorp-unet-2000-lr2e-ab & Eugeoter/artiwaifu-diffusion-1.0                          \\
                            & Lykon/AbsoluteReality                                                 & yuanbit/max-15-1e-6-1500                             & ehristoforu/Visionix-alpha                                \\ \bottomrule
\end{tabular}
}
\label{tab:name_model}

\end{table*}

\section{The Architecture of WFE}
\label{sec:arc_wfe}
In Sec.~\ref{sec:feature_extraction}, we have introduced the Watermark Feature Extractor (WFE). Here, we provide details of its architecture, as shown in Tab.~\ref{tab:wfe_architecture}. 
For the WFE, we use a 12-layer CNN with convolutional and fully connected layers, ReLU activations, and layer normalization, outputting a 100-dimensional feature vector. 
The WFE processes input images with a resolution of $256 \times 256$ through a series of $3 \times 3$ convolutional layers with stride 2, progressively reducing the spatial dimensions to an $8 \times 8$ feature map. Each convolutional layer is followed by a ReLU activation to introduce non-linearity. The resulting feature map is flattened and passed through two dense layers: the first projects it to a hidden dimension of 512, stabilized by LayerNorm, and the second produces a 100-dimensional watermark feature vector. A sigmoid activation function is applied to the output, ensuring that the values are in the range [0, 1], suitable for representing watermark features.

Training uses SGD optimizer with a learning rate of 0.01, momentum of 0.9, and a scheduler with a 0.5 decay factor every 50 steps.
\begin{table}[t]
\centering
\caption{Detailed architecture of the Watermark Feature Extractor. The table lists the parameters for each layer, including input channels, output channels, kernel size, stride, and activation function.}
\label{tab:wfe_architecture}
    \resizebox{\columnwidth}{!}{% 
\begin{tabular}{cccccc}
\toprule
\textbf{Layer} & \textbf{Type}       & \textbf{Input Channels} & \textbf{Output Channels} & \textbf{Kernel Size} & \textbf{Stride} \\ 
\midrule
1              & Conv2D              & 3                       & 32                       & $3 \times 3$         & 2               \\ 
2              & ReLU                & -                       & -                        & -                   & -               \\ 
3              & Conv2D              & 32                      & 32                       & $3 \times 3$         & 1               \\ 
4              & ReLU                & -                       & -                        & -                   & -               \\ 
5              & Conv2D              & 32                      & 64                       & $3 \times 3$         & 2               \\ 
6              & ReLU                & -                       & -                        & -                   & -               \\ 
7              & Conv2D              & 64                      & 64                       & $3 \times 3$         & 1               \\ 
8              & ReLU                & -                       & -                        & -                   & -               \\ 
9              & Conv2D              & 64                      & 64                       & $3 \times 3$         & 2               \\ 
10             & ReLU                & -                       & -                        & -                   & -               \\ 
11             & Conv2D              & 64                      & 128                      & $3 \times 3$         & 2               \\ 
12             & ReLU                & -                       & -                        & -                   & -               \\ 
13             & Conv2D              & 128                     & 128                      & $3 \times 3$         & 2               \\ 
14             & ReLU                & -                       & -                        & -                   & -               \\ 
15             & Flatten             & -                       & -                        & -                   & -               \\ 
16             & Dense               & -                       & 512                      & -                   & -               \\ 
17             & ReLU                & -                       & -                        & -                   & -               \\ 
18             & LayerNorm           & -                       & 512                      & -                   & -               \\ 
19             & Dense               & -                       & 100                      & -                   & -               \\ 
20             & Sigmoid             & -                       & -                        & -                   & -               \\ 
\bottomrule
\end{tabular}
}
\end{table}
}

 \bibliographystyle{IEEEtran}

\bibliography{main}

\newpage

\section{Biography Section}
\vspace{-33pt}

\begin{IEEEbiography}[{\includegraphics[width=1in,height=1.25in,clip,keepaspectratio]{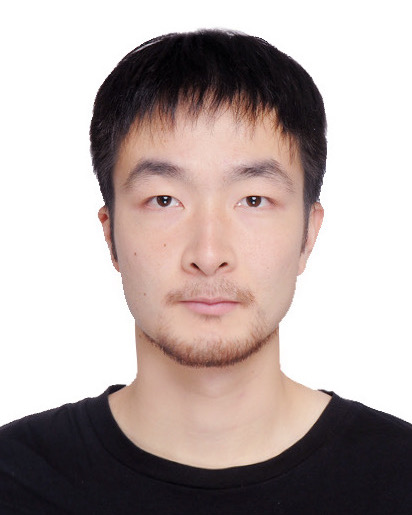}}]{Zhonghao Yang}
received the BEng degree from Northeastern University (P. R. China). He is currently pursuing the Ph.D. degree at East China Normal University. His research interests include  AI security and network and systems security.
\end{IEEEbiography}

\vspace{-33pt}

\begin{IEEEbiography}[{\includegraphics[width=1in,height=1.25in,clip,keepaspectratio]{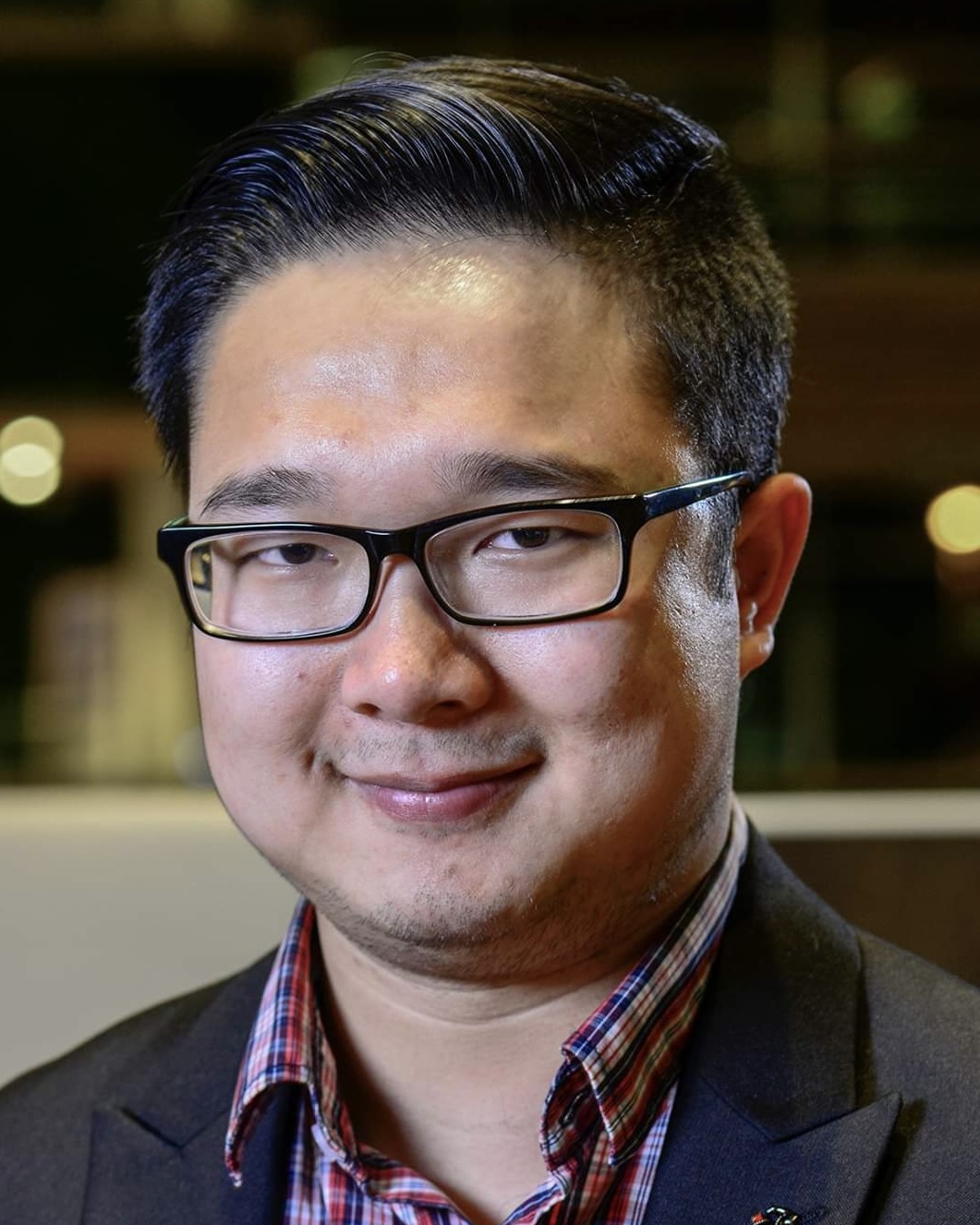}}]{Linye Lyu} is currently a Ph.D. student major in Computer Science and Technology in Harbin Institute of Technology, Shenzhen. His research interest are the AI safety issues in real-world scenarios such as autonomous driving and large languange models. He has published papers at leading conferences and journals like NeurIPS, ICML and TDSC. He obtained two master degrees in Artificial Intelgience and Electronics and ICT Engineering from Katholieke Universiteit Leuven (KU Leuven) in 2021 and 2017, respectively. He obtained his bachelor’s degree from  KU Leuven and University
of Electronic Science and Technology of China (UESTC) in 2016.
\end{IEEEbiography}
\vspace{-33pt}

\begin{IEEEbiography}[{\includegraphics[width=1in,height=1.25in,clip,keepaspectratio]{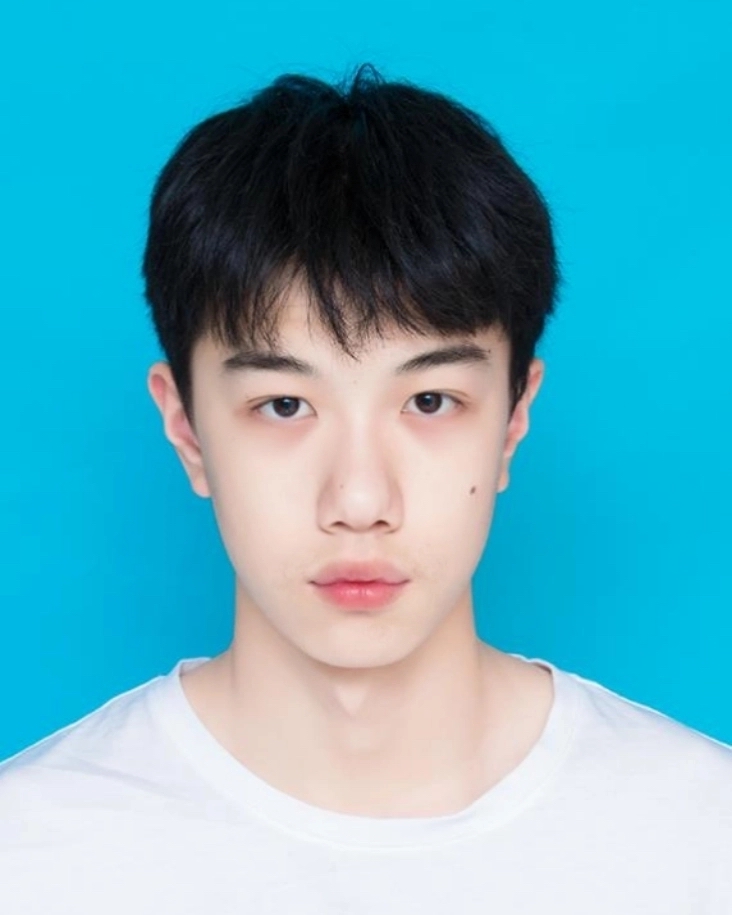}}]{Xuanhang Chang} is currently a master's student majoring in computer science at Harbin Institute of Technology, Shenzhen. His main research field is AI security, and he is currently working on digital image watermarking and watermarking for multimodal models. He obtained his bachelor's degree from Harbin Institute of Technology.
\end{IEEEbiography}
\vspace{-33pt}

\begin{IEEEbiography}[{\includegraphics[width=1in,height=1.25in,clip,keepaspectratio]{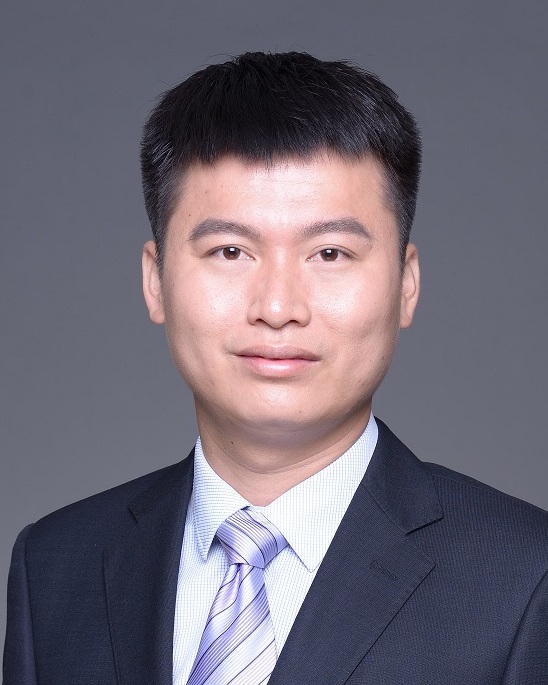}}]{Daojing He} received the B. Eng.(2007) and M. Eng. (2009) degrees from Harbin Institute of Technology (China) and the Ph.D. degree (2012) from Zhejiang University (China), all in Computer Science. He is currently a professor in the School of Computer Science and Technology, Harbin Institute of Technology, Shenzhen, China. His research interests include network and systems security. He is on the editorial board of some international journals such as IEEE Communications Magazine.
\end{IEEEbiography}

\vspace{-33pt}
\begin{IEEEbiography}[{\includegraphics[width=1in,height=1.25in,clip,keepaspectratio]{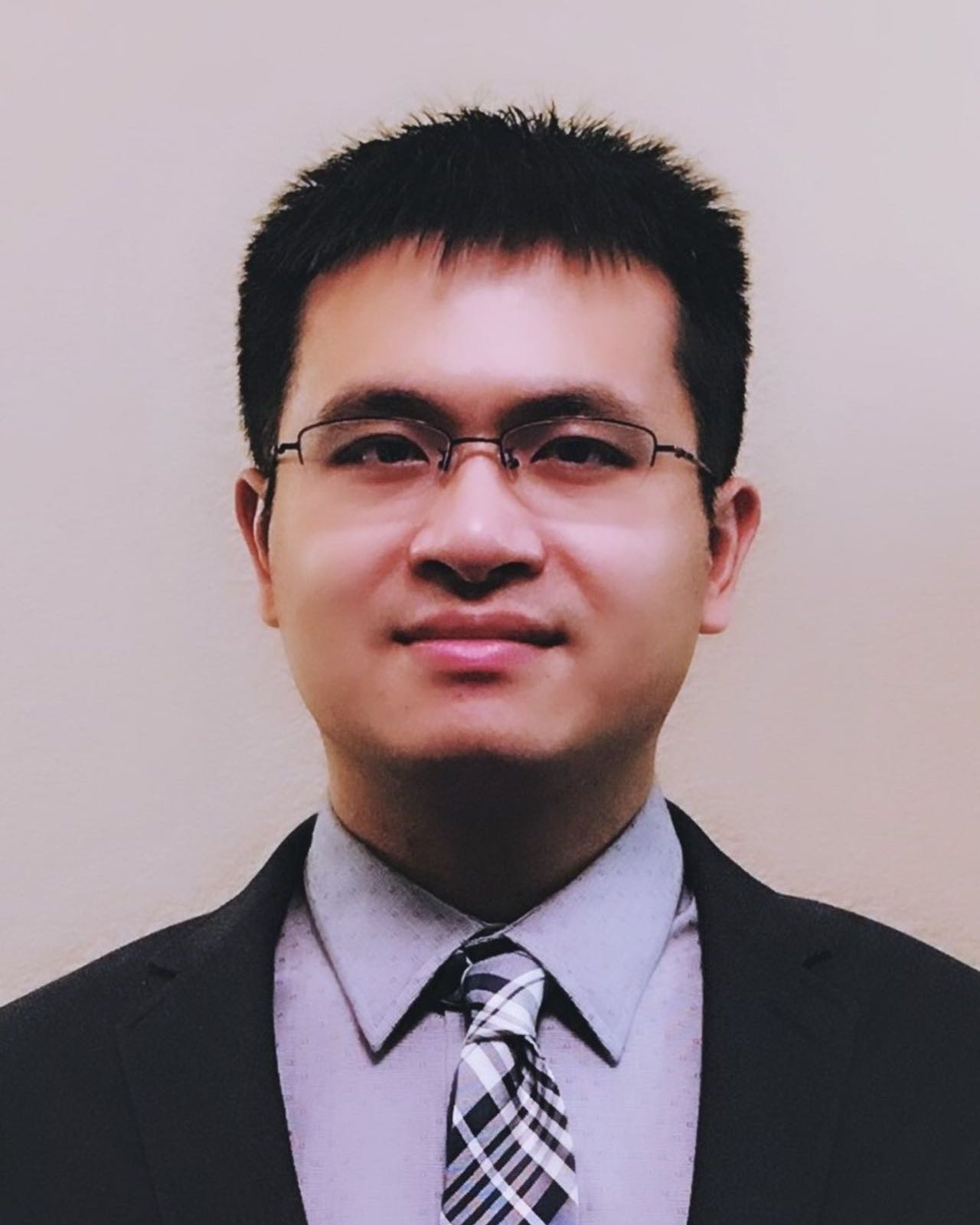}}]{Cheng Zhuo}
(Senior Member, IEEE) received his B.S. and M.S. from Zhejiang University, Hangzhou, China, in 2005 and 2007. He received his Ph.D. from the University of Michigan, Ann Arbor, in 2010. 
Dr. Zhuo is currently Qiushi Distinguished Professor at Zhejiang University, where his research focuses on hardware intelligence, machine learning-assisted EDA, and low power designs. He has authored/coauthored more than 200 technical papers, receiving 5 Best Paper Awards and 6 Best Paper Nominations. He is the recipient of ACM/SIGDA Meritorious Service Award and Technical Leadership Award, JSPS Faculty Invitation Fellowship, Humboldt Research Fellowship, etc. 
Dr. Zhuo has served on the organization/technical program committees of many international conferences, as the area editor for Journal of CAD\&CG, and as Associate Editor for IEEE TCAD, ACM TODAES, and Elsevier Integration. He is IEEE CEDA Distinguished Lecturer, a senior member of IEEE, and a Fellow of IET.
\end{IEEEbiography}

\vspace{-33pt}
\begin{IEEEbiography}[{\includegraphics[width=1in,height=1.25in,clip,keepaspectratio]{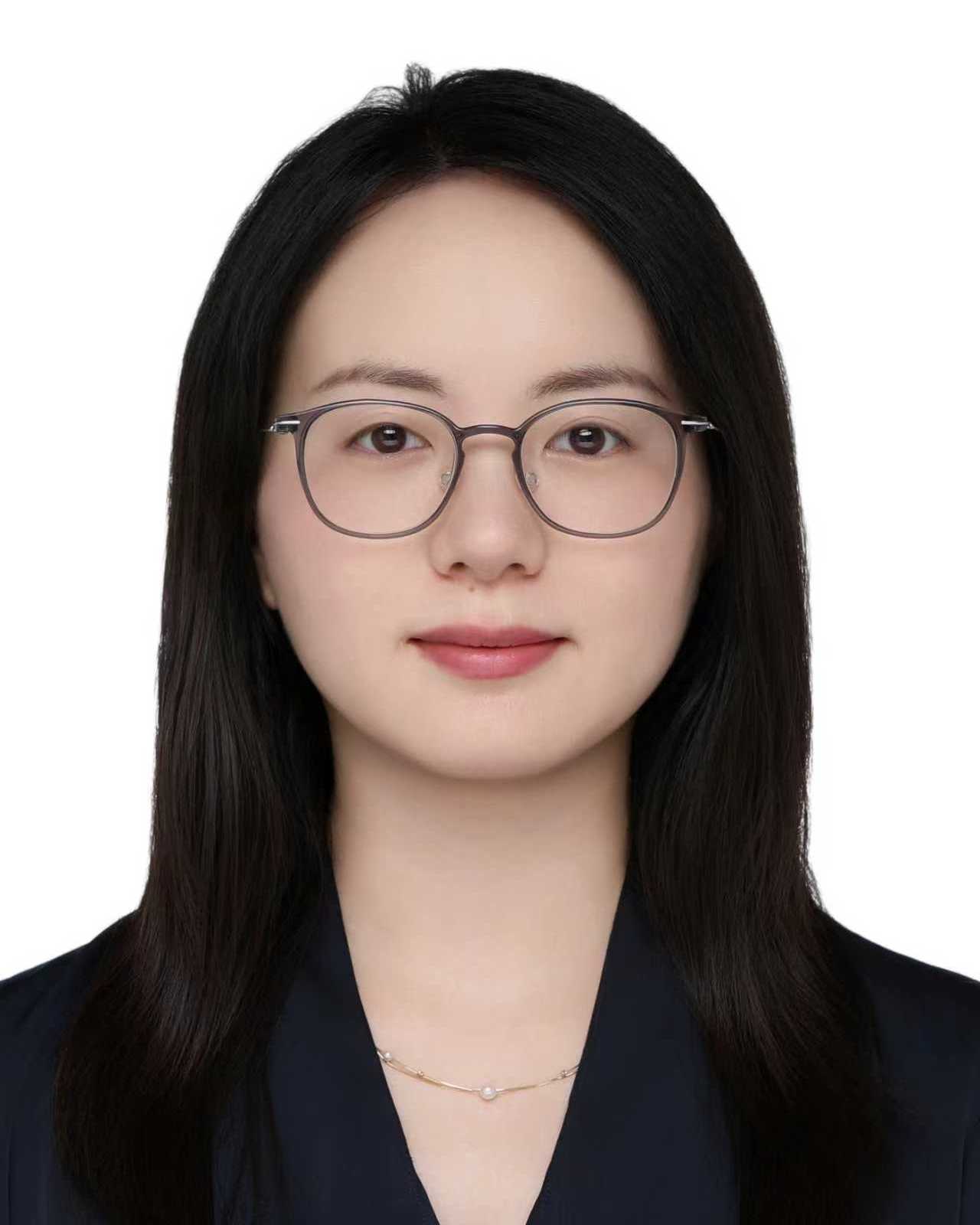}}]{Yu Li} is currently serving as a ZJU100 Professor at the School of Integrated Circuits, Zhejiang University. She received her Ph.D. degree from the Department of Computer Science and Engineering at the Chinese University of Hong Kong in 2022, obtained her master’s degree from the Katholieke Universiteit Leuven (KU Leuven) in 2017, and obtained her bachelor’s degree from the KU Leuven and the University
of Electronic Science and Technology (UESTC) in
2016. Li Yu’s main research direction is AI security
and testing. She has published many papers at leading conferences like CCS, NDSS, NeurIPS, ICML, and ISSTA.
She was nominated as a candidate for the 2022 Young Scholar Ph.D. Dissertation Award at The Chinese University of Hong Kong. In recognition of her outstanding work, she received the Best Ph.D. Dissertation Award at the 2022 Asian Test Symposium and the runner-up of the E. J. McCluskey Ph.D. Dissertation Award by the IEEE Test Technology Technical Council (TTTC).
\end{IEEEbiography}


\vfill

\end{document}